\documentclass[a4paper,fleqn,usenatbib]{mnras}
\usepackage{graphicx}
\usepackage{mathtools}
\usepackage{mathptmx}
\usepackage{txfonts}
\usepackage[T1]{fontenc}
\usepackage{ae,aecompl}
\usepackage{upgreek}
\usepackage{aas_macros}
\usepackage{lmodern}
\usepackage{subfigure}
\usepackage{hyperref}

\title[The clustering of post-starburst galaxies]{From starburst to quiescence: post-starburst galaxies and their large-scale clustering over cosmic time}
\author[A. Wilkinson et. al.]
{Aaron Wilkinson$^{1,2,3}$\thanks{E-mail: aaron.wilkinson91@gmail.com (AW)},
Omar Almaini$^{2}$, Vivienne Wild$^{3}$, David Maltby$^{2}$,
\newauthor{William G. Hartley$^{4}$, Chris Simpson$^5$, Kate Rowlands$^{6,7}$}\\
$^{1}$Sterrenkundig Observatorium, Universiteit Gent, Krijgslaan 281 S9, B-9000 Gent, Belgium\\
$^{2}$School of Department of Physics and Astronomy, University of Nottingham, University Park, Nottingham, NG7 2RD, UK\\
$^{3}$SUPA\thanks{Scottish Universities Physics Alliance}, School of Physics and Astronomy, University of St Andrews, North Haugh, St Andrews, KY16 9SS, U.K.\\
$^{4}$Institute for Particle Physics and Astrophysics, ETH Zurich, Wolfgang-Pauli-Strasse 27, CH-8093 Z{\"u}rich, Switzerland\\
$^{5}$Gemini Observatory, NSF's Optical-Infrared Astronomy Research Laboratory, 670 North A`oh\={o}k\={u} Place, Hilo, HI 96720, USA\\
$^{6}$Space Telescope Science Institute, 3700 San Martin Dr, Baltimore, MD 21218, USA\\
$^{7}$Johns Hopkins University, Department of Physics and Astronomy, Baltimore, MD 21218, USA
}
\begin{document}

\date{Accepted yyyy mm dd. Received yyyy mm dd; in original form yyyy mm dd}

\pagerange{\pageref{firstpage}--\pageref{lastpage}} \pubyear{0000}

\maketitle

\label{firstpage}

\begin{abstract}

We present the first study of the large-scale clustering of post-starburst (PSB) galaxies in the high redshift Universe ($0.5<z<3.0$). We select $\sim4000$ PSB galaxies photometrically, the largest high-redshift sample of this kind, from two deep large-scale near-infrared surveys: the UKIDSS Ultra Deep Survey (UDS) DR11 and the Cosmic Evolution Survey (COSMOS). Using angular cross-correlation techniques, we estimate the halo masses for this large sample of PSB galaxies and compare them with quiescent and star-forming galaxies selected in the same fields. We find that low-mass, low-redshift ($0.5<z<1.0$) PSB galaxies preferentially reside in very high-mass dark matter halos (M$_{\text{halo}}>10^{14}$\,M$_{\odot}$), suggesting they are likely to be infalling satellite galaxies in cluster-like environments. High-mass PSB galaxies are more weakly clustered at low redshifts, but they reside in higher mass haloes with increasing look-back time, suggesting strong redshift-dependent halo downsizing. These key results are consistent with previous results suggesting that two main channels are responsible for the rapid quenching of galaxies. While high-redshift ($z>1$) galaxies appear to be quenched by secular feedback mechanisms, processes associated with dense environments are likely to be the key driver of rapid quenching in the low-redshift Universe ($z<1$). Finally, we show that the clustering of photometrically selected PSBs are consistent with them being direct descendants of highly dust-enshrouded sub-millimetre galaxies (SMGs), providing tantalising evidence for the oft-speculated evolutionary pathway from starburst to quiescence.

\end{abstract}



\section{Introduction}
\label{sec:intro}

One of the major hurdles in extragalactic astronomy is understanding the origin of bimodality in the galaxy population observed today. While we have charted the build up of the quiescent red-sequence population in the aging Universe with a plethora of large-area multi-wavelength surveys, the mechanisms that cause the cessation of star formation in galaxies remain poorly understood. A number of galaxy properties have been found to correlate with both stellar mass of the galaxy and its environment. High-mass galaxies are typically bulge-dominated and red in colour \cite[e.g.,][]{vanderWel2008,Bamford2009}. These galaxies are also found in high-density environments such as clusters, particularly at low redshifts \citep[e.g.,][]{Kauffmann2003,Balogh2004}. One therefore asks the question: is galaxy evolution driven by external (environment) or internal (secular) factors? This widely debated dilemma is often referred to as the fundamental ‘nature’ versus ‘nurture’ problem.

The question of which factors determine the evolution of galaxy populations may be addressed by studying galaxies in transition. Some of these transition galaxies, post-starburst (PSB) galaxies (also often known as ``k$+$a'' or ``E$+$A'' galaxies), are identified by characteristic strong Balmer absorption lines \citep{Dressler1983, Wild2009} and a lack of nebular emission or ultraviolet flux, which indicates the excess of A-type stars and a deficiency of O- and B-type stars, respectively. The spectral features observed in PSB galaxies, in addition to their star formation histories \citep[SFHs][]{Wild2020}, demonstrate that these galaxies have undergone a rapid termination of star formation, \citep{Dressler1999, Poggianti1999, Goto2004, Dressler2004}, having formed a significant fraction of their stellar mass in the last Gyr \citep{Norton2001, Yang2004, Kaviraj2007, Wild2020}. PSB galaxies are relatively short-lived ($<1$\,Gyr), while their rarity is emphasised by their observed fractions of up to $\sim3\%$ of the entire galaxy population \citep{Goto2003, Wild2009, Wild2016, Alatalo2016a, Rowlands2018a}.

There are a variety of mechanisms that may initiate the post-starburst phase in galaxies. Many studies suggest that PSB galaxies are a product of gas-rich major mergers, which then undergo a starburst or ULIRG phase, before quenching adruptly \citep{Yang2004, Goto2004, Wild2009}. \citet{Maltby2019} found a prevalence of high velocity galactic-scale outflows and the apparent lack of AGN signatures in these high-redshift ($z\sim1$) PSBs, which may indicate quenching via stellar feedback induced by major mergers. This major merger scenario is also supported by the various galaxy evolution models proposed to explain the morphological transformation and the quenching of star formation in galaxies \cite[e.g.,][]{DiMatteo2005, Croton2006, Dekel2006, Hopkins2006, Somerville2008,Martig2009, Trayford2016}. Some models invoke a compaction scenario where inflows of cold gas feeding extended discs in galaxies cause these disks to become unstable and contract \citep{Dekel2009, Zolotov2015}.

Models also suggest AGN feedback is responsible for the removal of residual gas and dust at the end of the merger-induced starburst \citep[e.g.,][]{Silk1998, Hopkins2005,Narayanan2008, Trayford2016} and for the maintenance of the host galaxy's quiescence, by keeping the gas sufficiently heated \citep{Best2006}. Indeed, the presence of AGN have been detected in PSB galaxies, particularly at low redshift \citep[e.g.,][]{Yan2006,Wild2007,Alatalo2011, Cicone2014}. However, recent studies of the gas content in PSB galaxies have revealed the presence of substantial gas reservoirs, that remain even 500\,Myr to 1\,Gyr after the starburst \citep{Zwaan2013, Rowlands2015, French2015,Alatalo2016b,Suess2017}, suggesting that while AGN (and stellar) feedback reduces the star formation rates of these galaxies, full quiescence may not yet be achieved.

Galaxies can also be quenched by their host environments. PSB galaxies appear to be prevalent in clusters at redshifts $z<1$ \citep{Dressler2004, Tran2004, Poggianti2009, Socolovsky2018, Paccagnella2019}, although many have also been observed in field environments \citep[e.g.,][]{Zabludoff1996, Quintero2004, Pawlik2018}. Quenching processes associated with dense environments include starvation \citep{Larson1980}, strangulation \citep{Larson1980}, thermal evaporation \citep{Cowie1977} and ram-pressure stripping \citep{Gunn1972}. It has been suggested that the latter process, which involves the hydrodynamical interactions between the galaxy gas and the hot rarefied intracluster medium, is the mechanism responsible for the rapid and efficient quenching in PSB galaxies in cluster environments \citep{Paccagnella2017, Owers2019, Vulcani2020}.

Our understanding of low-redshift PSB galaxies have been complemented by the recently developing census of their high-redshift counterparts. So far, studies have shown that various properties of PSB galaxies depend on redshift: $z>1$ PSB galaxies are shown to be extremely compact \citep{Yano2016,Almaini2017,Maltby2018,Wu2018}, with high Sersic indices \citep{Almaini2017,Maltby2018} and stellar mass functions similar to those of quiescent galaxies \citep{Wild2016}. In contrast, $0.5<z<1$ PSB galaxies were found to have lower Sersic indices \citep{Maltby2018} and stellar mass functions that are consistent with star-forming galaxies \citep{Wild2016}.

Complementary to the various high-redshift analyses of PSB galaxies is the study of the properties of their host dark matter halos in which galaxies transition from the blue cloud to the red sequence. Galaxy clustering analyses quantify the spatial distribution of galaxies within the large-scale structure and provide a powerful method to constrain halo masses. There are a number of techniques that can be used to study the clustering of a galaxy population, the 2-point correlation function being the most widely used of these. It is well established that passive galaxies cluster more strongly than their star-forming counterparts \citep{Meneux2006, Williams2009, Hartley2010, Furusawa2011, Wake2011, Jullo2012, Hartley2013, Wilkinson2017,Cowley2019} and preferentially reside in more massive dark matter halos (M$_{\text{halo}}>5\times10^{12}$M$_{\odot}$). By comparing the clustering of red-sequence and star-forming populations, these studies suggest that the halo mass scale at which galaxies transition is around M$_{\text{halo}}\sim10^{12.5}$\,M$_{\odot}$. Investigating the clustering of galaxies caught in transition will therefore either confirm this halo mass scale or reveal deeper complexities of the halo-galaxy connection.

In this work, we extend previous studies to include the angular clustering of photometrically-selected PSB galaxies, and investigate their connections to the more common galaxy populations. We make use of a cross-correlation technique to statistically associate the sample of PSB galaxies to much larger $K$-band selected galaxy samples in the UKIDSS Ultra Deep Survey (UDS) and the Cosmic Evolution Survey (COSMOS). This analysis will allow us to study the dark matter halos inhabited by PSB galaxies for the first time, and investigate the evolutionary link between star-forming galaxies and those on the red sequence, across cosmic time.

The structure of this paper is as follows: Section 2 contains the discussion of our 
data sets and the selection of PSB galaxies with a PCA method; Section 3 describes our clustering methodology; in Section 4 we show the results of the clustering analysis; in Section 5, we discuss the implications, aided by a comparison to other galaxy populations, and a more general connection between galaxy properties and their clustering, ending with our 
conclusions and further work in Section 6. Throughout 
this paper we assume a $\lambda$-CDM cosmology with 
$\Omega_{\text{M}}=0.3$, $\Omega_{\lambda}=0.7$, H$_0=70$\,kms$^{-1}$Mpc$^{-1}$ and $\sigma_8=0.9$. All magnitudes 
are given in the AB system, unless otherwise stated.
\vspace{-0.3cm}

\section{DATA AND SAMPLE SELECTION}
\label{sec:data}

In this section, we describe the two deep large-area fields employed in this work: the UKIRT Infrared Deep Sky Survey \citep[UKIDSS,][]{Lawrence2007} Ultra Deep Survey (UDS) Data Release 11 (DR11, Almaini et al., in prep), and the Cosmic Evolution Survey \citep[COSMOS,][]{Scoville2007}. As we shall discuss, the deeper depth of the UDS complements the wider field of COSMOS, which allows us to accrue large samples of galaxies across a wide range of stellar masses. We will also discuss the process of using a principal component analysis \citep[PCA, using the methodology of][]{Wild2014} in selecting the key galaxy populations in deep photometric surveys, including PSB galaxies, and the comparison samples of red-sequence and star-forming galaxies.

\vspace{-0.3cm}

\subsection{UKIDSS UDS}
We use $K$-band selected samples from the UDS, complemented by matching multi-wavelength 
photometric data. Covering 0.8 square degrees, the UDS is an ultra deep survey in the $J, H$
and $K$ wavebands. Reaching a $5\sigma$ limiting depth of $J=25.6, H=25.1$, and $K=25.3\,$mag in $2''$ diameter apertures, the final UDS data release (DR11) is the deepest near-infrared survey to date over such a large area (Almaini et al., in prep). 

In addition to the three aforementioned near-IR bands, the photometric auxiliary data includes the $B, V, R, i', z'$ optical bands from the Subaru XMM-Newton Deep Survey 
(SXDS, \citealt{Furusawa2008}), the CHFT Megacam $u$-band and three IR bands from the \textit{Spitzer} UDS Legacy Program (SpUDS, PI:Dunlop).
The latter program comprises data in channels 1 and 2 ($3.6$ and $4.5\,\mu$m, respectively) of the IR Array Camera (IRAC), as well as observations in the 5.8, $8.0\,\mu$m and the MIPS $24\,\mu$m wavebands. The latter three wavebands are not utilised in this work however, as they have poor FWHM ($>2$\,arcsec) and relatively shallow limiting magnitudes ($\sim22.3$ at 5$\sigma$). In addition to SpUDS, we include complementary deeper IRAC data from the Spitzer Extended Deep Survey \citep[SEDS;][]{Ashby2013}. Finally, we include deep $Y$-band observations from the VISTA VIDEO survey \citep{Jarvis2013}, with a $5\sigma$ limiting depth of $J=24.4\,$mag. The coincident area covered by all the aforementioned surveys is 0.62 square degrees, after removing bright stars and image artefacts.

Photometric redshifts were computed using the methodology of \citet{Simpson2013}. Using the publicly available code \textsc{eazy} \citep{Brammer2008}, the photometry was fitted with a linear combination of six solar metallicity, simple stellar population templates with ages spaced between 30\,Myr and 10\,Gyr, and additional younger templates that were dust-reddened using a Small Magellanic Cloud extinction law. In addition to extremely deep photometry, the UDS field contains a total of $\sim7000$ galaxies with secure spectroscopic redshifts, with most of the $z>1$ spectra obtained in the UDSz project \citep{Bradshaw2013,McLure2013}. This project was undertaken using a combination of the VIsible MultiObject Spectrograph (VIMOS) and FORS2 instruments on the European Southern Observatory Very Large Telescope (ESO Large Programme 180.A-0776, PI: Almaini). Comparing photometric and spectroscopic redshifts, the normalised median absolute deviation - the uncertainty in the distribution of d$z$/($1+z$) - was found to be $\sigma_{NMAD}=0.019$. Further details of the photometric redshifts determined for DR11 will be presented in Hartley et al. (in prep.).

For our clustering analysis, the full redshift probability density distribution (PDF) is employed, following the methodology of \citet{Wake2011} and \citet{Hartley2013}.
According to the redshift PDFs, a single galaxy can be represented in multiple redshift slices, where each entry is weighted by the integrated probability between the limits of a given redshift slice. The broader the redshift PDF (which is certainly the case for galaxies at higher redshifts), the more entries a galaxy will have, with each entry having a smaller weight and therefore a smaller contribution to the clustering measurements at a given redshift slice. We verify that employing single best-fit redshifts in the clustering analysis gives consistent results, though with larger uncertainties. We refer the reader to Section 3 for more details.


\subsection{COSMOS}
The combined COSMOS and UltraVISTA project \citep{Scoville2007, McCracken2012} is a survey offering a wealth of deep multi-wavelength photometry, observed over a large area of $\sim1.5$ square degrees (after masking out stars). While it is significantly larger than the UDS field, it is not as deep; the full field reaches a depth of $K_s=23.9$ (at $5\sigma$ in a 2\,arcsec diameter aperture). However, coincident with the COSMOS field, an area of 4 ``ultra-deep stripes'', totalling $\sim0.5$ square degrees, is observed at a much deeper depth of $K_s=24.6$. Note that the depth and the completeness are not the same across the entire COSMOS field. For the clustering analysis, an extensive uniform field is preferable to constrain the clustering measurements at large scales. Therefore, our analyses are restricted to the full field of COSMOS, at the shallower depth.

The most recently published dataset, ``COSMOS2015'' \citep{Laigle2016}, is a combination of X-ray ($Chandra$), near ultra-violet (NUV; GALEX), near-IR \citep[WIRCam and UltraVISTA,][]{McCracken2010, McCracken2012}, IR (MIPS/$Spitzer$) and far-IR (PACS/$Herschel$ and SPIRE/$Herschel$) data. Supplementary optical data is composed of 6 broad bands ($B, V, g, r, i, z^{++}$) in addition to 12 medium bands and 2 narrow bands, all taken with the Subaru Suprime-Cam \citep{Taniguchi2007, Taniguchi2015}. The NIR data is a combination of $YJHK_{s}$ data obtained with the VIRCAM instrument on the VISTA telescope as part of the UltraVISTA project. Finally, the mid-IR dataset of IRAC $3.6\mu$m, $4.5\mu$m, $5.8\mu$m and $8.0\mu$m is a combination of the SPLASH ($Spitzer$ Large Area Survey with Hyper-Suprime-Cam) COSMOS dataset, S-COSMOS \citep{Sanders2007}, the $Spitzer$ Extended Mission Deep Survey, the $Spitzer$-CANDELS and a number of other projects in COSMOS.

Readers are referred to \citet{Laigle2016} for an in-depth discussion of the publicly available, matching multi-wavelength catalogue and redshift estimates, so we briefly summarise the key points here. Objects in this field were selected by a $\chi^{2}$ sum of the combined $YJHK_{s}$ and $z^{++}$ images. \citet{Laigle2016} computed photometric redshifts with \textsc{LePhare} \citep{Arnouts2002, Ilbert2006}, using templates of spiral and elliptical galaxies from \citet{Polletta2007}, as well as templates of young star-forming galaxies from \citet{Bruzual2003} models (BC03) for fitting. The uncertainty in the distribution of d$z$/($1+z$) is $\sigma_{NMAD}=0.01$, derived using the large variety of spectroscopic surveys in COSMOS \citep[e.g., zCOSMOS,][]{Lilly2007}. For the clustering analysis in COSMOS, we employ the full probability redshift distribution, in the same way as described for the UDS.




\subsection{Classifying galaxies: the Principal Component Analysis}
\label{sec:sampleselect}

Following the PCA methodology of \citet{Wild2014, Wild2016} with the UDS DR8, we calculate ``supercolours'', by applying a PCA technique to optical--near-infrared observed-frame photometric SEDs in the UDS DR11 and COSMOS fields, using the publicly available code from \citet{Wild2014}. The procedure involves optimising a linear combination of three eigenspectra, in which their corresponding amplitudes, supercolours (SCs), provide a compact representation of a wide variety of SED shapes, enabling us to classify four key galaxy populations. Based on where galaxies lie in the SC parameter space, we can separate a tight red-sequence from star-forming galaxies, as well as identify unusual populations: dusty star-forming galaxies with high SFRs, and PSB galaxies, which are selected as recently and rapidly quenched galaxies, having formed a significant fraction ($>10\%$) of their mass in the preceding $\sim$Gyr. For the clustering analysis, we combine the dusty and star-forming samples into one class (so that the ``dusty'' galaxies are henceforth identified as ``star-forming'').

The key changes made to the PCA technique introduced in \citet{Wild2014} are the extended ranges of redshift and rest-frame wavelength (from $\lambda=3000$\,\AA$-15000$\,\AA$~$ to $\lambda=2500$\,\AA$-15000$\,\AA), over which the eigenbasis is defined. Lowering the rest-frame wavelength limit improves our ability to classify galaxies at redshifts $z\sim2$, due to the inclusion of $z$-band rest-frame measurements of the blue end of the SEDs at this redshift. The resulting eigenbases are slightly different to those initially determined by \citet{Wild2014}, so that SCs defined in UDS DR8 are not directly equivalent to the SCs determined in UDS DR11.  In addition, we have a spectroscopic catalogue that has greatly increased in sample size, thanks to the combination of the UDSz project, the VANDELS project \citep{McLure2018,Pentericci2018} and, presented in \citet{Maltby2016}, the targeted VIMOS observations of PSB candidates (ESO programme 094.A-0410, PI: Almaini). Having a much larger sample of spectroscopically identified PSB galaxies has greatly improved our ability to identify PSB galaxies photometrically with the PCA. Therefore the original boundaries for classifying galaxies in UDS DR8 are recalibrated accordingly. Furthermore, the COSMOS and UDS filter sets are slightly different, so the boundaries constructed in UDS SC space must be transformed to equivalent boundaries that can classify galaxies in COSMOS in a consistent manner.

The classification boundaries in UDS SC space have been reconstructed in an optimal way to maximise both the accuracy and completeness rates of the photometric selection of PSB galaxies, based on the measurements of the equivalent widths of [O \textsc{ii}] and $\text{H}\delta$, and the strength of the 4000\,\AA$~$break of the spectroscopic sample. In previous literature \citep[e.g.,][]{Goto2007}, PSBs are selected by their equivalent widths in H$\delta$ ($W_{\text{H}\delta}$), such that $W_{\text{H}\delta}> 5$\,\AA. Subjecting the spectroscopic sample to various cuts (using only secure $z_{spec}$ and $W_{\text{H}\delta}$ measurements, $z_{spec}<1.4$ and S/N $>5$), we find that $\sim72\%$ (20/28) of photometrically-selected PSB candidates exhibit strong PSB spectral features, with a completeness rate of $\sim60\%$, consistent with those determined by \citet{Maltby2016}.

Some PSBs exhibit [O \textsc{ii}] emission lines, which may indicate the presence of AGN and/or residual star formation \citep{Yan2006}. If we reject PSB galaxies using the criterion of \citet{Tran2003}, $W_{[\text{O} \textsc{ii}]} < -5$\,\AA, the resulting PSB accuracy and completeness rates are $\sim64\%$ and $\sim70\%$, respectively\footnote{We note that the procedure and criteria for classifying PSB galaxies in the UDS field differ from those used in \citet{Wild2020}, where they employ the spectral fitting package \textsc{bagpipes} \citep{Carnall2018} to derive SFHs. Moreover, they showed that many PSB galaxies that are identifed spectroscopically but not photometrically (i.e., by their SCs) tend not to have PSB-like SFHs, with slowly declining SFRs. Conversely, photometrically identified PSB galaxies that are labelled as something else spectroscopically, have PSB-like SFHs, for which they have weak bursts of star formation followed by rapid truncation. These trends in SFHs emphasise the success in identifying PSB galaxies by photometric means.}. Finally, using measurements of $W_{[\text{O} \textsc{ii}]}$ and the strength of the 4000\,\AA$~$break, D4000, we recalibrate the quiescent/star-forming boundary to where the number of galaxies were evenly split between the star-forming ($W_{[\text{O} \textsc{ii}]} < -10$\,\AA) and quiescent samples (D4000$>$1.4 and $W_{\text{H}\delta}< 5$\,\AA). We plot the SC1 and SC2 values of the spectroscopically classified galaxies in the UDS and the recalibrated boundaries in Figure 1. We also include those that are classified as PSB photometrically but are not assigned to any population class spectroscopically (black triangles). Most of these unconfirmed PSB galaxies have $\text{H}\delta$ equivalent widths of 4\,\AA $<W_{\text{H}\delta}< 5$\,\AA, indicating that they have weak PSB features.

\begin{figure}
\centering
\begin{subfigure}
\centering
\includegraphics[height=0.43\textwidth]{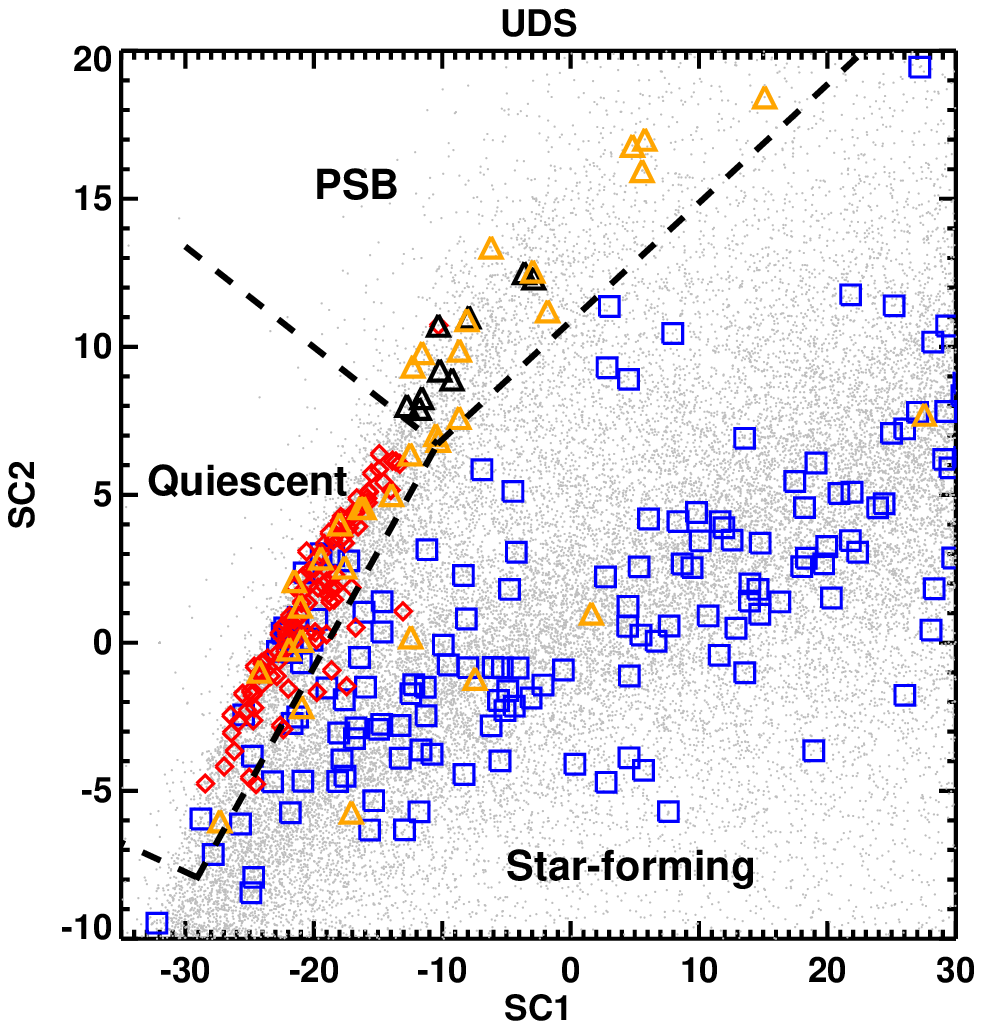}
\end{subfigure}

\begin{subfigure}
\centering
\includegraphics[height=0.43\textwidth]{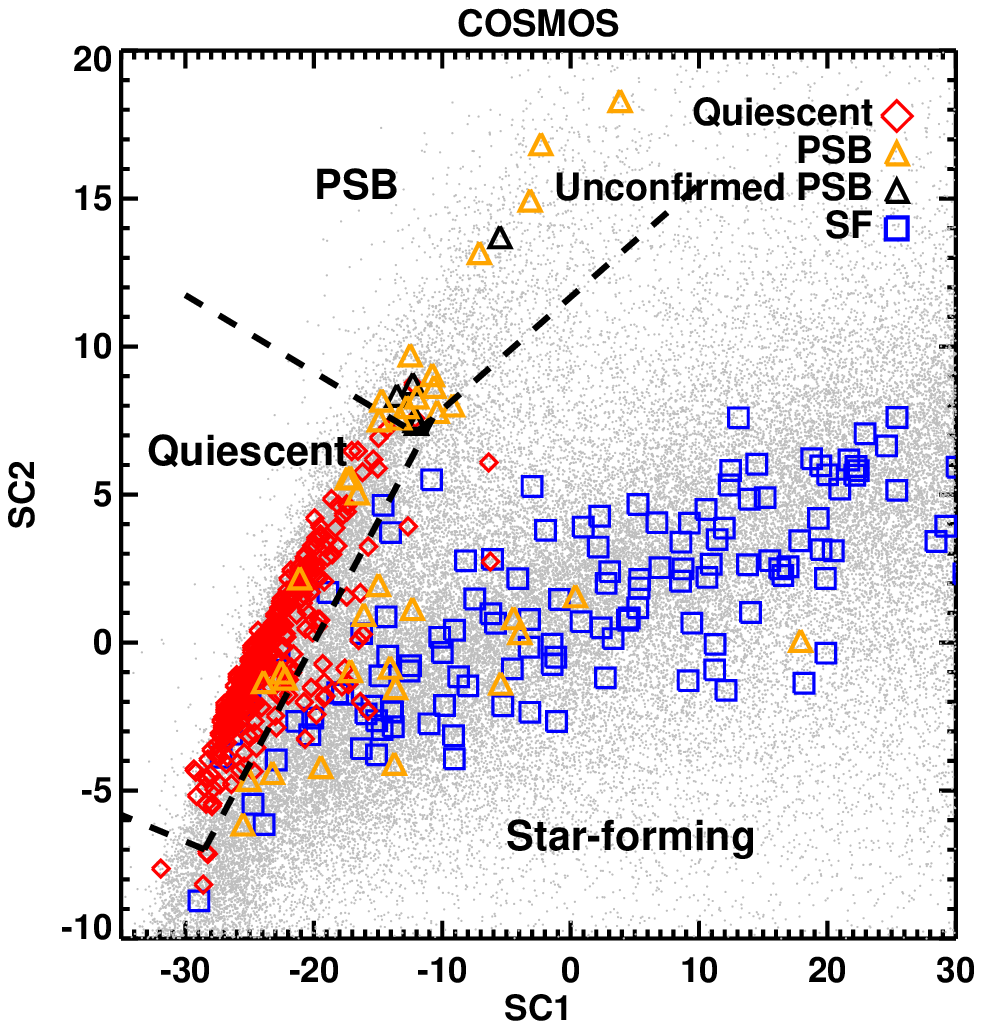}
\end{subfigure}
\caption{Top panel: A supercolour diagram zoomed into the PSB region, displaying the subset of galaxies with spectroscopy plotted over photometrically-selected galaxies (grey points) in the UDS.  The spectroscopic samples are divided into different populations as follows: PSB (orange, $W_{\text{H}\delta}> 5$\,\AA, $W_{[\text{O} \textsc{ii}]} > -5$\,\AA), star forming (blue, $W_{[\text{O} \textsc{ii}]} < -5$\,\AA) and red-sequence (red, D$4000 > 1.4$, $W_{[\text{O} \textsc{ii}]} > -5$\,\AA). Unconfirmed photometrically-selected PSB galaxies (i.e., those that fail to satisfy the criteria for each of the spectroscopic galaxy classification) are plotted as black triangles. The black dotted lines are boundaries which separate galaxies into different classes, placed in an optimal way to maximise accuracy and completeness rates. Bottom panel: The supercolour diagram for galaxies in the COSMOS surveys, where all symbols have the same meaning as described above.
\label{fig:halomass}}
\end{figure}

We next discuss the transformation of the UDS classification boundaries to those equivalent for COSMOS. We first construct mock SEDs that lie along the UDS SC1-SC2 demarcation lines in step sizes of $\Delta$SC1$=0.1$. For each value of SC1-SC2 we use the mean value of SC3 from the library of SEDs used to build the eigenbasis. We verify that shifting the average SC3 values slightly does not significantly impact the transformation, as expected, since the third eigenvector accounts for only $0.15\%$ of the variance of model SED shapes. The reconstructed SEDs are then projected onto the COSMOS eigenvectors to derive the equivalent SCs, where the boundary is drawn as a line of best fit.

We assess the robustness of the COSMOS classification boundaries, using the final data release of the zCOSMOS spectroscopic dataset \citep{Lilly2007}. We impose a cut of S/N $>5$, calculate the equivalent widths in the H$\delta$ and [O \textsc{ii}] spectral features, and classify galaxies based on these features. An accuracy rate of $\sim74\%$ (14/19 photometrically identified PSBs) was found for zCOSMOS. Filtering out possible star-forming galaxies with the cut $W_{[\text{O} \textsc{ii}]} < -5$\,\AA, the accuracy and completeness rates are $\sim69\%$ and $\sim70\%$, respectively. We confirm that the transformed quiescent/star-forming boundary lies along the region where $50\%$ of galaxies are spectroscopically classified as red-sequence galaxies (using the D4000 diagnostic), with the other $50\%$ being classified as star-forming. The spectral confirmation of PSB galaxies in the COSMOS and UDS fields is in excellent agreement, indicating that the transformation of the classification boundaries between the different datasets is robust.

We further evaluate the efficacy of the PCA as a function of redshift, which is largely determined by the availability of survey broad-band filters at a given redshift \citep{Wild2014}. The library of model SEDs, on which the PCA was applied, were sampled across all redshifts between 0.5 and 3.0 (or 2.5 for COSMOS), in every UDS/COSMOS broad-band filter, with their ``true'' SCs computed \citep[see][for details]{Wild2014}. Following the methodology of \citet{Wild2014}, we then produce a mock catalogue, by randomly assigning each model SED a single redshift in the range $0.5 < z < 3.0$ ($2.5$ for COSMOS) and calculate their ``mock'' SCs. The creation of the mock also takes into account the redshift inaccuracies and unknown normalisation. We classify the model SEDs based on both their true and mock SCs (using the classification boundaries described earlier), and calculate the following: 1) the PSB incompleteness, the fraction of true PSB SEDs that are no longer classified as PSB in the mock, and 2) the contamination rate, the fraction of mock PSBs that were originally classified as non-PSB. We find the PSB classification has incompleteness and contamination rates of $<15\%$ at all redshifts and masses, reducing to $<5\%$ at $0.8<z<1.2.$

While the effects of redshift and filter availability on the success of the PSB classification can be quantified, determining the relationship between the PCA efficacy and stellar mass (thus by extension the quality of the photometric measurements) is more complex to model and therefore beyond the scope of this work. We instead examine how the formal $1\sigma$ errors on the SCs derived for the UDS and COSMOS galaxy catalogues vary with the galaxy's stellar mass (details on the derivation of stellar mass are to be discussed shortly). These $1\sigma$ errors are propagated from the errors on the fluxes through the PCA algorithm. For high-mass log(M$^*/$M$_{\odot})>10$ PSB galaxies, the $1\sigma$ errors on both SC1 and SC2 are mostly $\sigma_{SC}=1.0$ at redshifts $0.5<z<1.0$. These errors typically multiply by $\sim2-2.5$ for the lowest mass (log[M$^*/$M$_{\odot}]=9$) PSBs examined in this work.

Combining the effects of both mass and redshift together, we are confident that the PSB classification, as a whole, is effective. Even though the efficacy of our classifications is reduced for the lowest-mass PSB galaxies, as well as for PSB galaxies found at the highest redshift, our results and conclusions of the clustering analysis remain robust. Indeed, as we will present in Section 4, low-mass PSB galaxies are the most strongly clustered population of all. Any significant contamination would reduce the large clustering amplitudes of this population to values consistent with those of the quiescent or star-forming galaxies, both of which have lower clustering amplitudes.

Finally, we calculate various galaxy properties of interest. Stellar masses, SFRs and sSFRs were computed using a Bayesian analysis outlined in \citet{Wild2016}, which accounts for the degeneracy between physical parameters. The library of tens of thousands of \citet{Bruzual2003} models were fitted to the supercolours derived for the observed galaxies and a probability density function (PDF) was obtained for each physical property. The value of each property is extracted from the median of the PDF, with $1\sigma$ errors determined from the 16th and 84th percentiles of the PDF. As described earlier, for the purposes of our clustering analysis presented in Section 3, each galaxy may be represented and weighted in multiple redshift intervals, based on their redshift PDFs. Supercolours (and hence galaxy classifications), stellar masses, SFRs and sSFRs are therefore recomputed for each entry, at the minimum $\chi^2$ redshift within the target redshift interval. A single galaxy may therefore have different classifications across multiple redshift intervals.


\subsection{The final sample selection}

Having selected the PSB, red-sequence and star-forming galaxies with the PCA, we use various quality and completeness cuts to clean up these samples. We first discuss our final galaxy samples in the UDS. For each of the galaxy populations, we impose cuts of $K<24.5\,$mag and $\chi^2<30$; the latter quantity is from photometric redshift fitting. We sort the samples into redshift intervals of size $\Delta(z)=0.5$, from $z=0.5$ to $z=3.0$. We also apply $90\%$ mass completeness limits on the samples, as a function of redshift and galaxy class (star-forming, red-sequence and PSB). These redshift-dependent limits were computed using the prescription of \citet{Pozzetti2010}, for which we fit with second order polynomials of mass limits (see Table 1). We plot the redshift-mass diagram for the PSB sample in Figure 2, showing the mass completeness limits of each galaxy population, and the chosen redshift and stellar mass intervals. We also report the subsample sizes and their sum of weights (determined by their redshift probability distributions; see Section 2.1) in Table 2 and Table 3 (at the end of this paper).

\begin{figure}
\centering
\begin{subfigure}
\centering
\includegraphics[height=0.44\textwidth]{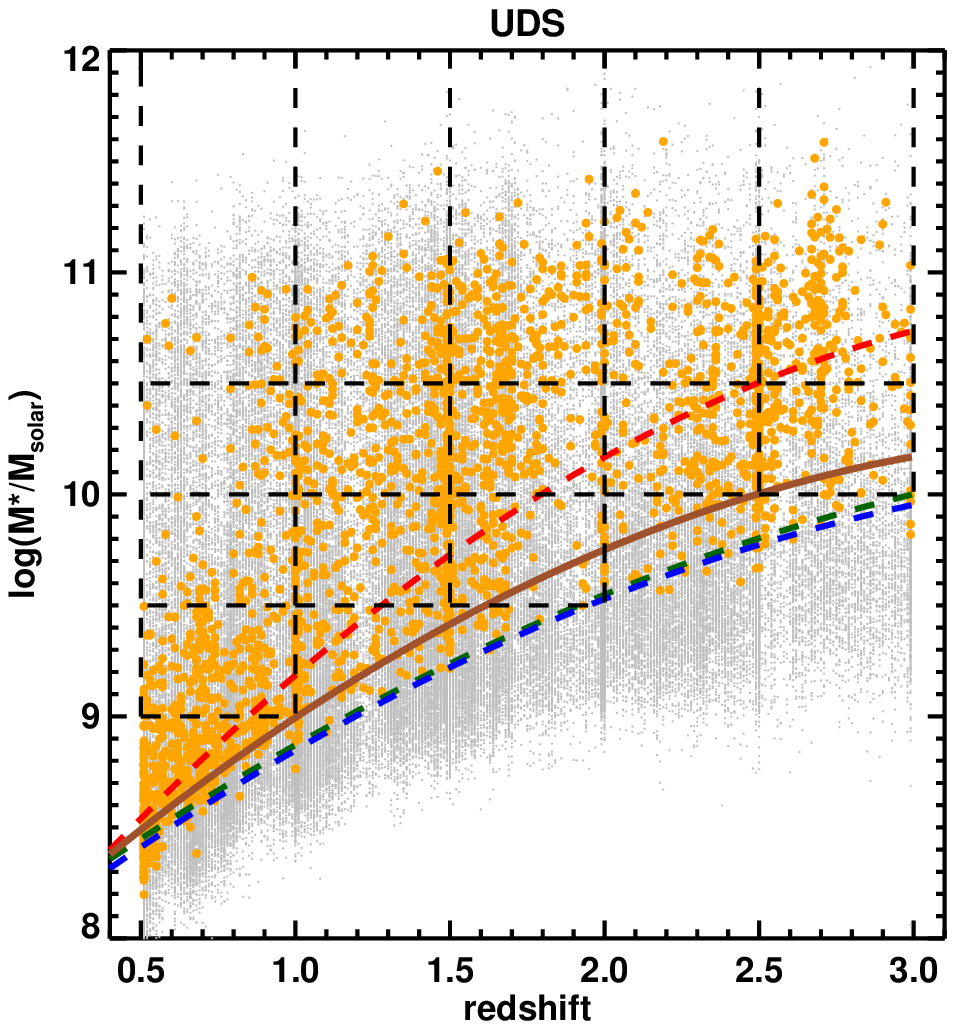}
\end{subfigure}
\begin{subfigure}
\centering
\includegraphics[height=0.44\textwidth]{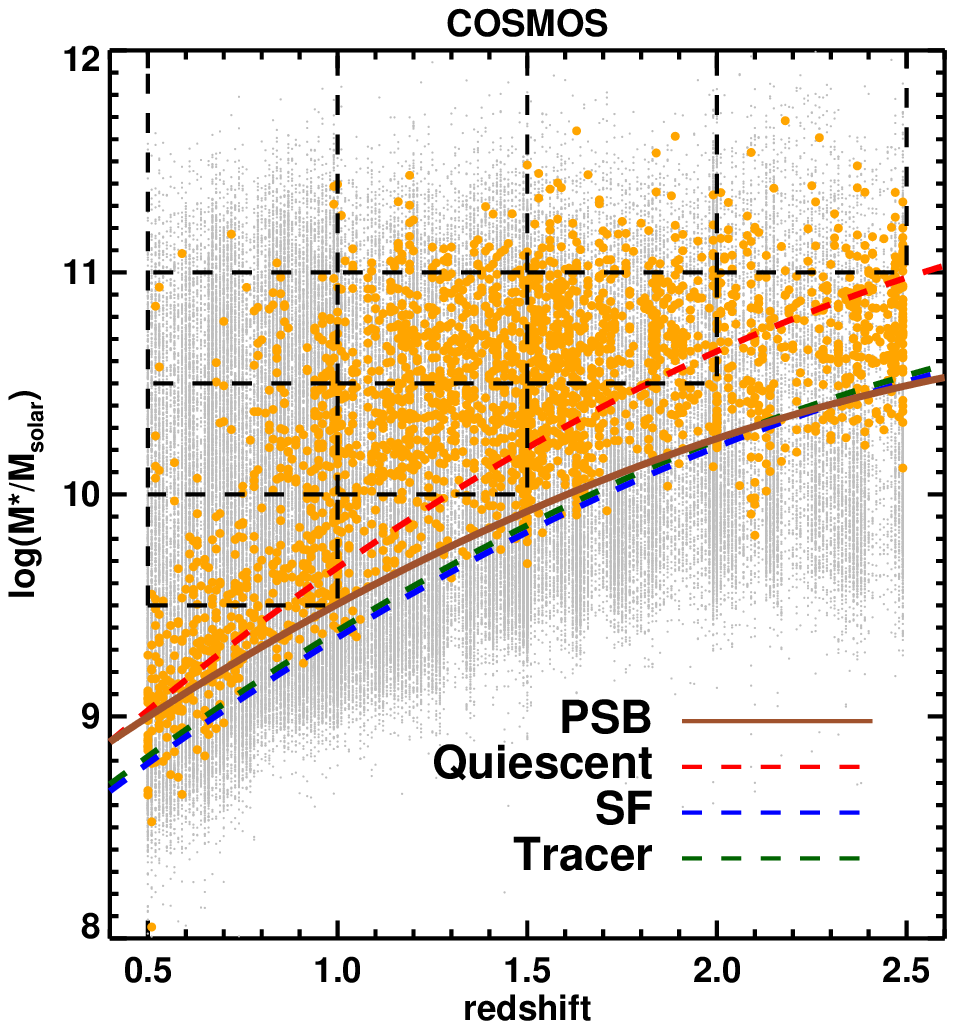}
\end{subfigure}
\caption{Top panel: Stellar mass versus photometric redshifts for the UDS (top) and COSMOS (bottom) galaxy samples. The PCA-selected PSB galaxies are overplotted in orange. The solid brown line shows the second-order polynomial fit to the $90\%$ mass completeness limits computed for the PSB sample. Dashed red, blue and green lines display the $90\%$ mass completeness limits for quiescent, star-forming and combined samples, respectively. The black dashed lines indicate the redshift and stellar mass bins used for the clustering analysis.
\label{fig:halomass}}
\end{figure}

\begin{table}
\centering
\begin{tabular}{ c c c c }
\hline

Class & $a$ & $b$ & $c$\\
\hline
\textbf{UDS} \\
Star-forming & -0.13 & 1.07 & 7.91\\
Quiescent & -0.21 & 1.60 & 7.81\\
PSB & -0.17 & 1.26 & 7.95\\
Tracer & -0.11 & 1.01 & 7.97\\
\hline
\textbf{COSMOS} \\
Star-forming & -0.18 & 1.41 & 8.13\\
Quiescent & -0.21 & 1.60 & 8.28\\
PSB & -0.18 & 1.29 & 8.40\\
Tracer & -0.18 & 1.41 & 8.16\\
\hline
\end{tabular}
\caption{Table of UDS and COSMOS $90\%$ mass completeness limit polynomials for each class, in the form log(M$^*/$M$_{\odot})=az^{2}+bz+c$.}
\label{table:1}
\end{table}

In a similar fashion to the UDS, a number of cuts were imposed to clean our subsamples in the full field of COSMOS: a cut of $K_s<23.4$ was applied to the galaxy samples, shallower than the $K$-band cut for the UDS. In the publicly available COSMOS2015 catalogue \citep{Laigle2016}, $\chi^2$ values from photometric redshift fitting were provided in the reduced form. Hence, we impose a reduced $\chi^2=2.5$ on the samples, roughly equivalent to the $\chi^2$ cut used in the UDS. In principle, the PCA can classify galaxies out to $z=3$ in COSMOS, however, the $90\%$ mass completeness limit at this redshift is log(M$^*/$M$_{\odot})\sim11$, which rejects the majority of galaxies. The completion of the ultra-deep photometric observations in the full field of COSMOS is required to investigate the clustering of PSB galaxies to $z=3$. Therefore, we restrict the analysis to $0.5<z<2.5$ for COSMOS. The SED type-specific $90\%$ mass completeness limits were imposed on all galaxy populations, in the same way as the UDS. We plot the mass completeness limits, and the chosen redshift and stellar mass intervals in the bottom panel of Figure 2.

Finally, for both fields, we combine all PSB, red-sequence and star-forming galaxies within the redshift and mass limits described above to build the volume-limited tracer galaxy population with which we cross-correlate each of our subsamples in the clustering analysis. In both of the UDS and COSMOS fields, we repeated the clustering analyses using the more conservative $95\%$ mass completeness limits for each of our galaxy populations (including the tracer population). We verify that the resulting clustering measurements are consistent with those presented in this paper (with the $90\%$ completeness limits), albeit with larger uncertainties.


\section{CLUSTERING ANALYSIS}
\label{sec:analysis}

There are a number of techniques that can be used to study the clustering of a galaxy population, the 2-point auto correlation function (ACF) being the most widely used of these \citep[see][for a detailed review]{Coil2013}. The ACF, which traces the amplitude of galaxy clustering as a function of scale, is defined by the excess number of pairs of galaxies above random, as a function of their separation \citep{Groth1977, Peebles1980}. To calculate the angular ACF, we employ the \citet{Landy1993} estimator, described by

\begin{equation}
w(\theta)=\frac{DD(\theta)-2DR(\theta)+RR(\theta)}{RR(\theta)},
\label{eq:LS}
\end{equation}

\noindent
where $DD(\theta)$, $DR(\theta)$ and $RR(\theta)$ are the galaxy-galaxy, galaxy-random and random-random normalised numbers of pairs, respectively. Because the UDS and COSMOS surveys are finite in size, the clustering calculations may be underestimated by a constant factor. The required correction, known as the integral constraint, becomes important on angular scales comparable to survey size, though it is negligible when much smaller scales are considered. We numerically estimate the integral constraint over the survey geometry, using the method of \citet{Roche1999},

\begin{equation}
C=\frac{\sum RR(\theta).w(\theta)}{\sum RR(\theta)}.
\label{eq:IC}
\end{equation}

Thus, the integral constraint depends on the intrinsic clustering of galaxies and we can estimate this constraint by adopting some form for 
$w(\theta)$, which we describe shortly. To correct the angular correlation function, this constant is added to all values of $w(\theta)$.

Galaxies form within dark matter halos and the correlation function of a galaxy sample is intimately related to the typical mass of those halos: the more clustered a galaxy population is, the more massive the dark matter halos which host them. This relationship comes about because massive dark matter halos are biased with respect to the overall dark matter distribution, forming from the densest peaks of the primordial dark matter distribution. The bias is a monotonically increasing function of redshift and halo mass \citep{Mo1996}; the bias is naturally higher at earlier epochs of galaxy formation, as the first galaxies to form will collapse in the most overdense regions of space. Hence, these galaxies are biased tracers of the dark matter distribution. With time, newly formed galaxies become unbiased tracers of the mass distribution, since they start to form in less dense regions of the large scale structure.

 Following the method of \citet{Hartley2013}, we therefore assume $w(\theta)$ in equation \ref{eq:IC} to be the 
angular correlation function of the underlying dark matter distribution traced by galaxy populations.
Using the formalism of \citet{Smith2003}, we calculate the non-linear power spectrum to determine the dark matter correlation function. A power spectrum in $k$-space can be Fourier transformed to produce a corresponding 2-point correlation function in real space. Using the galaxy redshift distribution, $n(z)$, we can project this correlation function to obtain the angular dark matter correlation function. Because the bias, $b$, intimately connects the observed galaxy clustering to the intrinsic clustering of dark matter, it is computed by the following relationship,

\begin{equation}
w_{\rm{obs}}(\theta)=b^2\times w_{\rm{dm}}(\theta),
\label{eq:wobs}
\end{equation}

\noindent
where $w_{\rm{obs}}$ and $w_{\rm{dm}}$ are the correlation functions of the observed galaxy population 
and dark matter distribution, respectively. We fit the model correlation function $w_{\rm{dm}} \times b^2$ to the observed correlation function $w_{\rm{obs}} + C \times b^2$, where the latter term is the integral constraint. This fitting is carried out by minimising the correlated $\chi^2$ statistic at large scales.

At large scales, $w_{\rm{dm}}$ and $w_{\rm{obs}}$ are dominated by the linear ($w_{\rm{linear}}$) correlation regime, where the dark matter ACFs are well described by linear gravity theory. The two regimes deviate at small scales however. Non-linear effects become much more significant at these scales, leading to a boost in the excess number of pairs in the correlation function. The assumption that the galaxy population traces the dark matter distribution is no longer valid. In other words, the excess contribution from $w_{\rm{non-linear}}$ becomes non-negligible. For the purposes of constraining our bias measurements from the correlation functions, we define the transition limit between the linear and non-linear scales, $w_{\rm{total}} < 3\times w_{\rm{linear}}$, where $w_{\rm{total}} = w_{\rm{linear}} + w_{\rm{non-linear}}$.\footnote{Readers may note that this factor of three times is different to the more widely used factor of two, which may give rise to the concern we are fitting correlation functions deeper into the non-linear regime, where our model assumptions could no longer be valid. We compared the bias measurements, their errors and the reduced $\chi^2$ from the fitting procedure, using both transition limits. We find that using the multiplication factor of three results in no systematic increases in the bias measurements, $\sim10\%$ smaller errors, and reduced $\chi^2$ values that are closer to 1, compared to using the factor of two. We therefore verify that our choice has little impact on the clustering measurements, while achieving smaller errors and better fits.} In addition to this lower limit, we adopt an upper limit of $\theta=0.4$ degrees, where separations in $\theta$ approach the size of our surveys. From the bias parameter, we estimate the halo masses of our samples using the \citet{Mo2002} formalism.

Our ability to derive accurate ACFs and well-constrained halo masses is significantly impaired by using populations of very small sample sizes. However, we can compute a closely related correlation function: the 2-point cross correlation function (CCF), using the positions of the much larger sample of galaxies tracing the large-scale structure. We cross-correlate the target sample population ($D_{\rm{s}}$) with a full volume-limited tracer population ($D_{\rm{t}}$), using the following estimator:

\begin{equation}
w(\theta)=\frac{D_{\rm{s}}D_{\rm{t}}(\theta)-D_{\rm{s}}R(\theta)-D_{\rm{t}}R(\theta)+RR(\theta)}{RR(\theta)}.
\label{eq:cross}
\vspace{0.15cm}
\end{equation}

\noindent
where each term is normalised by their respective total pair counts. The tracer population includes all PSB, quiescent and SF galaxies with masses above the $90\%$ mass completeness limit, for a given redshift slice. Cross-correlating a population of small sample sizes with a much more numerous tracer population that samples the dark matter distribution greatly boosts the number of pairs. Compared to the auto-correlation, the CCF achieves dramatically reduced statistical uncertainties. The pair counts produced by the CCF must also be weighted. Each galaxy's contribution to a redshift slice is weighted by the integral of its probability distribution between the limits of a given redshift slice (see Section 2.1). Hence, the contribution each pair makes to the CCF is the product of their respective objects' weights in a given redshift interval. We compute the ACF of the large tracer population and the CCF of the target sample crossed with the tracer to derive the bias of the tracer population ($b_{\rm{t}}$) and the CCF bias, $b_{\rm{st}}$ respectively, using equation~\ref{eq:wobs}. These two bias values can then be used to infer the bias of the target sample population, $b_{\rm{s}}$, as follows:

\begin{equation}
b_{\rm{s}}=\frac{b^2_{\rm{st}}}{b_{\rm{t}}}.
\label{eq:bias}
\end{equation}

We can extract the ACF of the target sample population by multiplying the CCF by $(b^2_{\rm{st}}/b^2_{\rm{t}})$, allowing us to compare the ACF of all sample populations. To calculate these correlation functions, we use random catalogues with sample sizes 10 times larger than that of the tracer population. To estimate the errors of the correlation functions, we employ a bootstrap analysis with 100 repetitions, resampling the galaxy population with replacement. We derive $w(\theta)$ for each of the 100 bootstrap samples and then calculate the covariance errors on $w(\theta)$, taking into account that the measurements of $w(\theta)$ at different separations are correlated with each other. As the robustness of the clustering analyses greatly depend on the sample sizes, a lower limit on the sample size is imposed, by rejecting subsamples that have less than 30 objects, after binning in redshift and stellar mass.


\section{RESULTS}
\label{sec:results}

\begin{figure*}
\centering
\includegraphics[height=1.15\textwidth, trim = 0 0 0 0,clip]{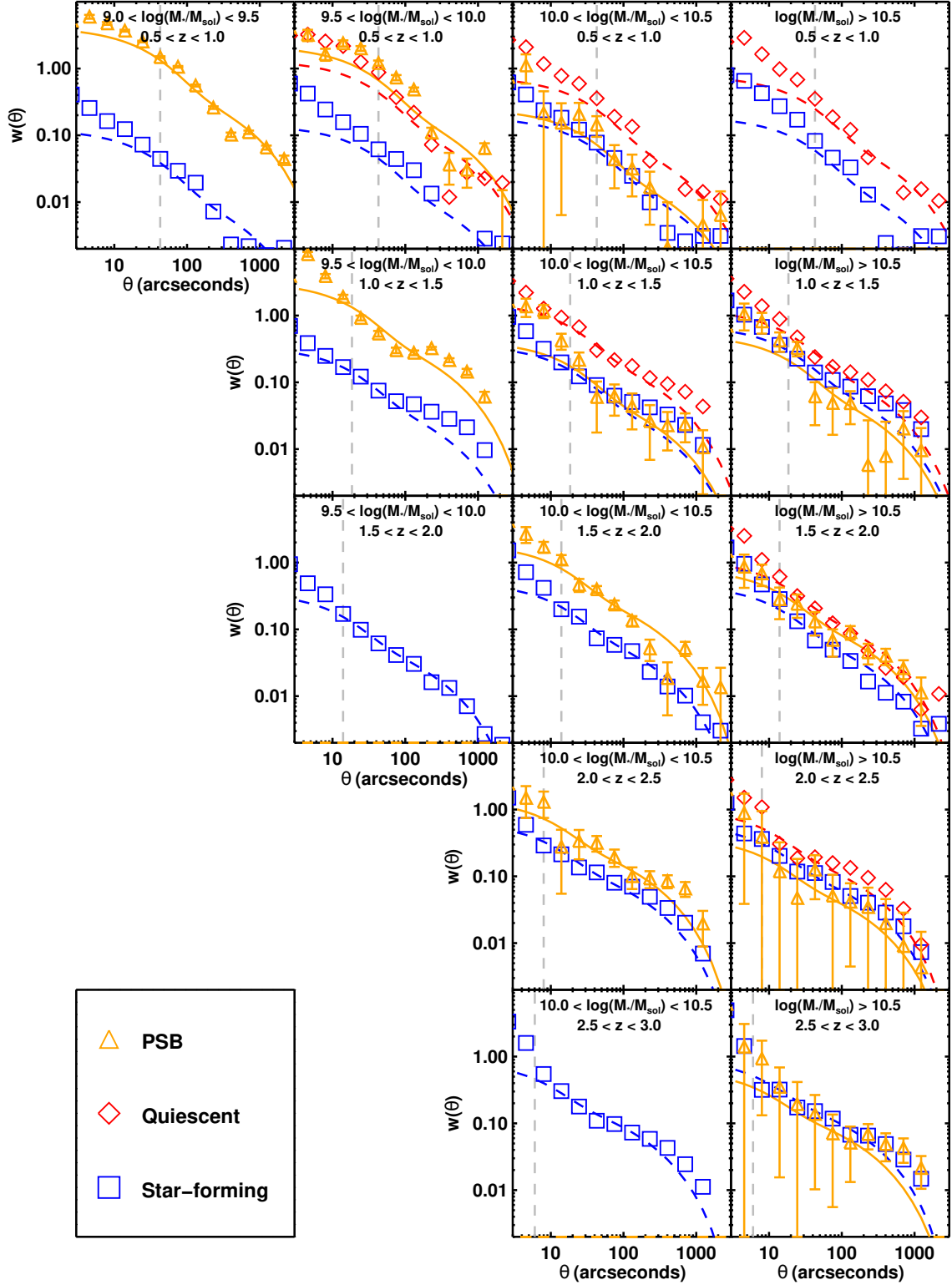}
\caption{The correlation functions for PSB (orange triangles), red-sequence (red diamonds) and star-forming (blue squares) galaxies in the UDS field, split by redshift and stellar mass intervals. The solid (dashed) lines are dark matter correlation functions fitted onto the observed PSB (red-sequence and star-forming) galaxy correlation functions. The vertical grey lines indicate the minimum separation $\theta$ that the models are fitted to the data, given by $w_{\rm{non-linear}} = 3\times w_{\rm{linear}}$ (see Section 3). Error bars for the red-sequence and star-forming galaxy correlation functions have been omitted for clarity.
\label{fig:halomass}}
\end{figure*}

\begin{figure*}
\centering
\includegraphics[height=0.95\textwidth, trim = 0 0 0 0,clip]{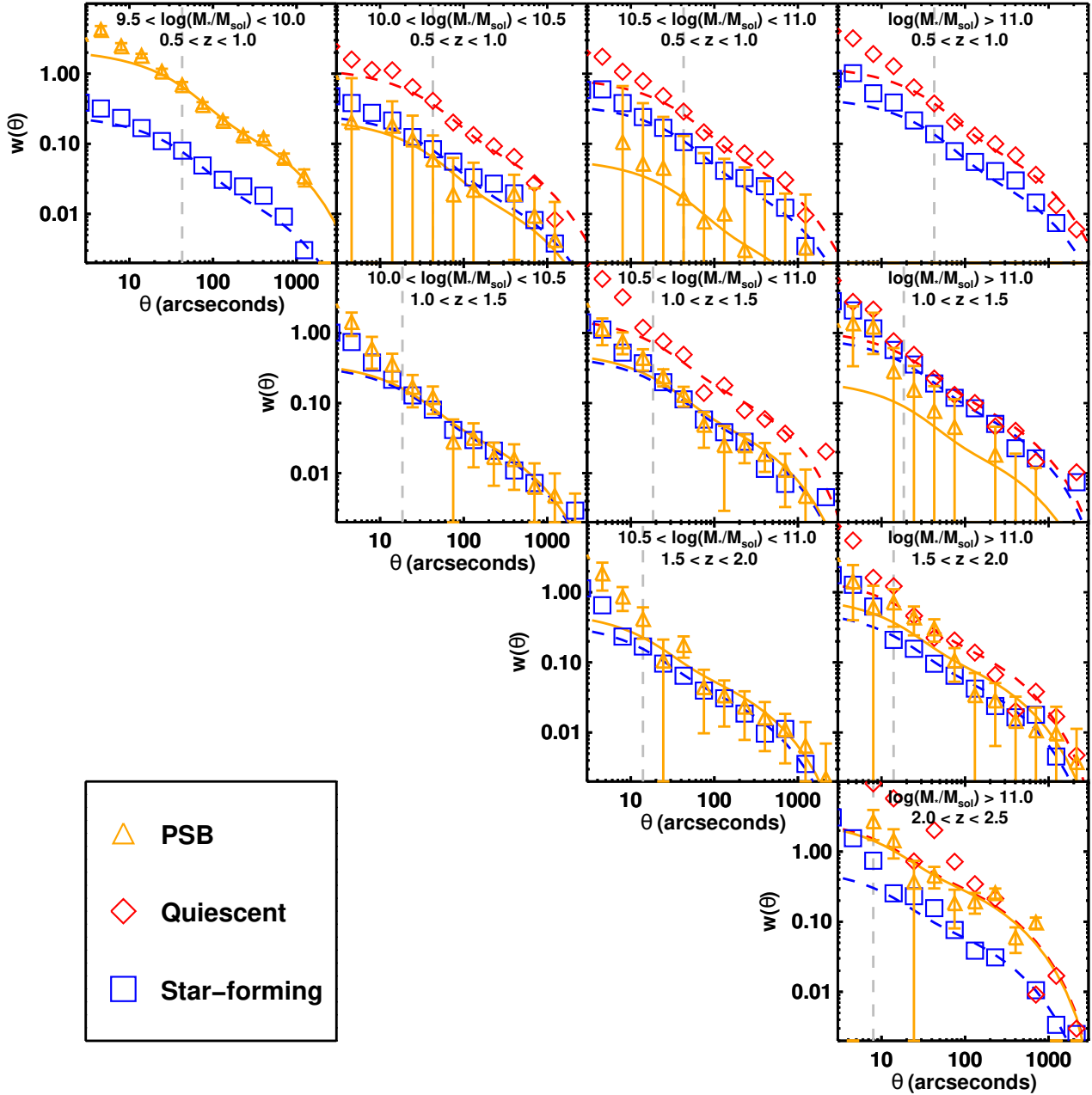}
\caption{The correlation functions for PSB, red-sequence and star-forming galaxies in the COSMOS field, separated by redshift and stellar mass intervals. All symbols and lines have the same meanings as presented in figure 3.
\label{fig:halomass}}
\end{figure*}

\begin{figure*}
\centering

\includegraphics[height=1.23\textwidth,trim = 0 0 0 0,clip]{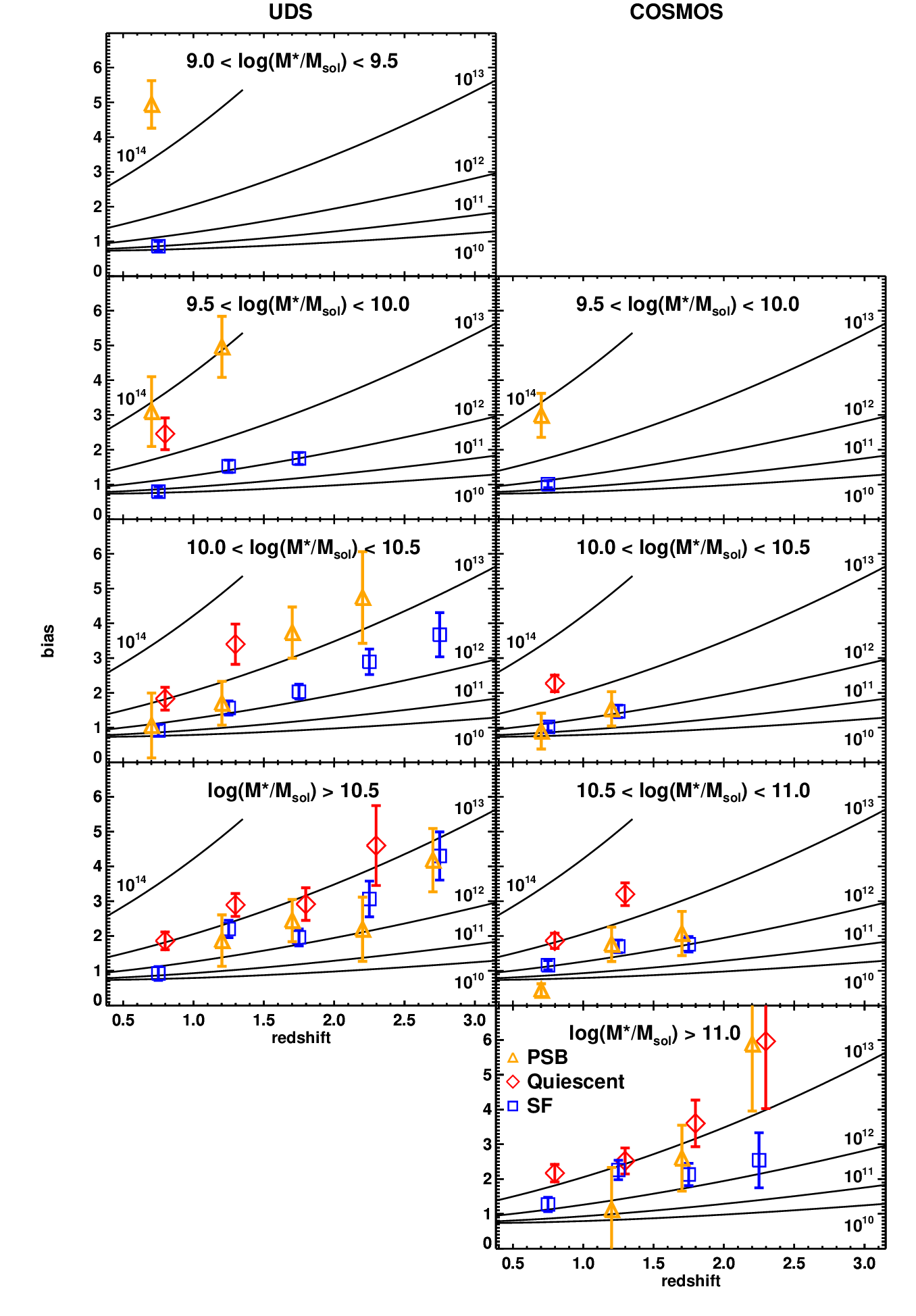}

\caption{Galaxy bias versus redshift for quiescent (red diamonds), star-forming (blue squares) and PSB galaxies (orange triangles) in different stellar mass intervals in the UDS (left) and COSMOS (right) fields. For clarity, the points are offset slightly, with respect to the $x$-axis. The solid lines show the evolution of bias for dark matter halos of varying mass (labelled, in solar masses). The bias of PSB galaxies, and therefore their inferred halo mass, depends strongly on mass and redshift, with low-mass and low-redshift PSB galaxies typically occupying high mass halos, and high-mass PSB galaxies occupying halos of higher mass with increasing redshift.
\label{fig:halomass}}
\end{figure*}

In this section, we present the clustering measurements of PSB galaxies as a function of redshift and stellar mass for the first time, and compare with the red-sequence and star-forming galaxies. 

\subsection{Correlation functions}

Applying the cross-correlation analysis outlined in Section 3, we plot the resulting correlation functions for the UDS and COSMOS fields, in Figure 3 and Figure 4, respectively. The auto-correlation functions are derived from their respective cross-correlation functions multiplied by $(b^2_{\rm{CCF}}/b^2_{\rm{tracer}})$. We find that red-sequence galaxies, on the whole, exhibit stronger clustering amplitudes with respect to their star-forming counterparts, at all redshifts and stellar masses considered. This is in excellent agreement with previous studies, including \citet{Hartley2013}, who also presented clustering measurements of red-sequence and star-forming galaxies in the UDS field, but selected with $UVJ$ criteria.

PSB galaxies exhibit a range of clustering signals, depending on the redshift and stellar mass considered. At a given redshift interval, lower mass PSB galaxies are more strongly clustered, while high-mass PSB galaxies are more weakly clustered and comparable to star-forming galaxies of the same mass. This relationship is most apparent at low redshifts ($0.5<z<1.5$). Considering the redshift dependence, the clustering results indicate that high-mass PSB galaxies, in most cases, have stronger clustering amplitudes with increasing redshift (with the exception of the UDS highest mass interval, where the relationship is more elusive). It remains to be seen, however, whether this relationship holds for low-mass galaxies, due to the limited completeness of our high-redshift galaxy samples and the depth of our large-area surveys.

\subsection{Halo masses}

We convert our projected angular correlation functions into equivalent bias measurements to describe the clustering in real space, using the methodology outlined in Section 3. We plot these bias measurements in Figure 5 and Figure 6, to demonstrate how the clustering of our galaxy samples depend on redshift and stellar mass, and how they relate to their inferred halo masses. Focussing our attention on the red-sequence and star-forming populations first, we confirm, for both fields, that red-sequence galaxies are more strongly clustered than their star-forming counterparts. The red-sequence population tends to occupy dark matter halos of mass M$_{\text{halo}}\sim10^{13}$M$_{\odot}$, more massive than halos (M$_{\text{halo}}\sim10^{12}$M$_{\odot}$) preferentially traced by star-forming galaxies, consistent with a wealth of previous studies \citep[e.g.][]{Meneux2006, Williams2009, Furusawa2011, Wake2011, Jullo2012, Hartley2013}. For both populations, we find weak dependence of clustering on stellar mass.

PSB galaxies appear to reside in halos of a range of masses. Various trends become clear when we deconstruct the sample by their stellar mass intervals (Figure 6). In both fields, the clustering of PSB galaxies depends on stellar mass: low-mass and low-redshift, $9.0<$log(M$^*/$M$_{\odot})<10.0$, $0.5<z<1.0$, PSB galaxies in the UDS reside in very high-mass halos (M$_{\text{halo}}\sim10^{14}$M$_{\odot}$), indicating that low-mass PSBs are prevalent in clusters. We see a similar result in COSMOS for PSB galaxies of mass $9.5<$log(M$^*/$M$_{\odot})<10.0$, albeit with slightly lower values of bias and halo masses. The existence of a known supercluster at $z\sim0.65$ in the UDS field \citep{vanBreukelen2006} may account for the boosted bias measurements in the lowest redshift interval. The most massive PSB galaxies are not as strongly clustered as red-sequence galaxies, tracing halos of mass M$_{\text{halo}}\sim10^{12}$M$_{\odot}$. There is also evidence for a clustering evolution of massive PSB galaxies: log(M$^*/$M$_{\odot})>10.5$ PSB galaxies occupy higher mass halos with increasing redshift, indicating halo downsizing.

To investigate the possibility of star-forming contaminants in the PSB selection (which may lead to low clustering amplitudes), we impose a stricter selection criteria, by raising the PSB SF/boundaries in the SC parameter space ($\Delta$SC2$=+2$). We confirm that the resulting PSB bias measurements are consistent with those presented in Table 2, albeit with larger uncertainties. Possible contamination from star-forming galaxies are therefore unlikely to cause the low bias measurements for high-mass PSB galaxies.


\section{DISCUSSION}

The clustering results have revealed interesting stellar mass- and redshift-dependent relationships between recently quenched galaxies and their host halos. We discuss the implications of these results by evaluating the possible processes that lead galaxies to quench their star formation abruptly within their halos, and their connections with other rare galaxy populations.

\subsection{Halo occupation effects}

We evaluate the relationship between a galaxy and its host halo, and speculate on satellite fractions of various galaxy populations. We first consider the stellar mass to halo mass (SMHM) ratio, which is defined as the galaxy's stellar mass divided by the mass of the galaxy's host dark matter halo. The SMHM ratio is effectively a quantitative measure of the conversion of baryons into stars. A key feature of most models \citep[e.g.][]{Croton2006,Zheng2007,Guo2010,Cen2011,Behroozi2013,Moster2013,Behroozi2019} is the so-called `pivot mass', a few $10^{12}$M$_{\odot}$, where halos of this mass are most efficient at converting baryons into stars. Beyond this mass, models predict that AGN feedback reduces the efficiency dramatically, with maximum quenching at $10^{13}$M$_{\odot}$. This feature induces a bending in the SMHM relation, which describes the strong correlation between the stellar mass of a galaxy and the mass of its host halo.

For our galaxy samples, we convert their bias measurements into halo masses using the formalism of \citet{Mo2002}, and plot them against their stellar masses in Figure 6. We note that the measured clustering biases calculated in this study are occupation-weighted biases. Hence, the higher the fraction of satellite galaxies (galaxies that are in the halo of another dominant central galaxy) in a given galaxy sample, the further the measured characteristic halo mass may be from the minimum halo mass of halos that host only central galaxies \citep{Hartley2013}. In other words, a departure from the predicted SMHM relation of central galaxies \citep[e.g.][as plotted in Figure 6]{Moster2013} may indicate higher satellite fractions in a galaxy sample.

\begin{figure*}
\centering

\begin{subfigure}
\centering
\includegraphics[height=0.58\textwidth]{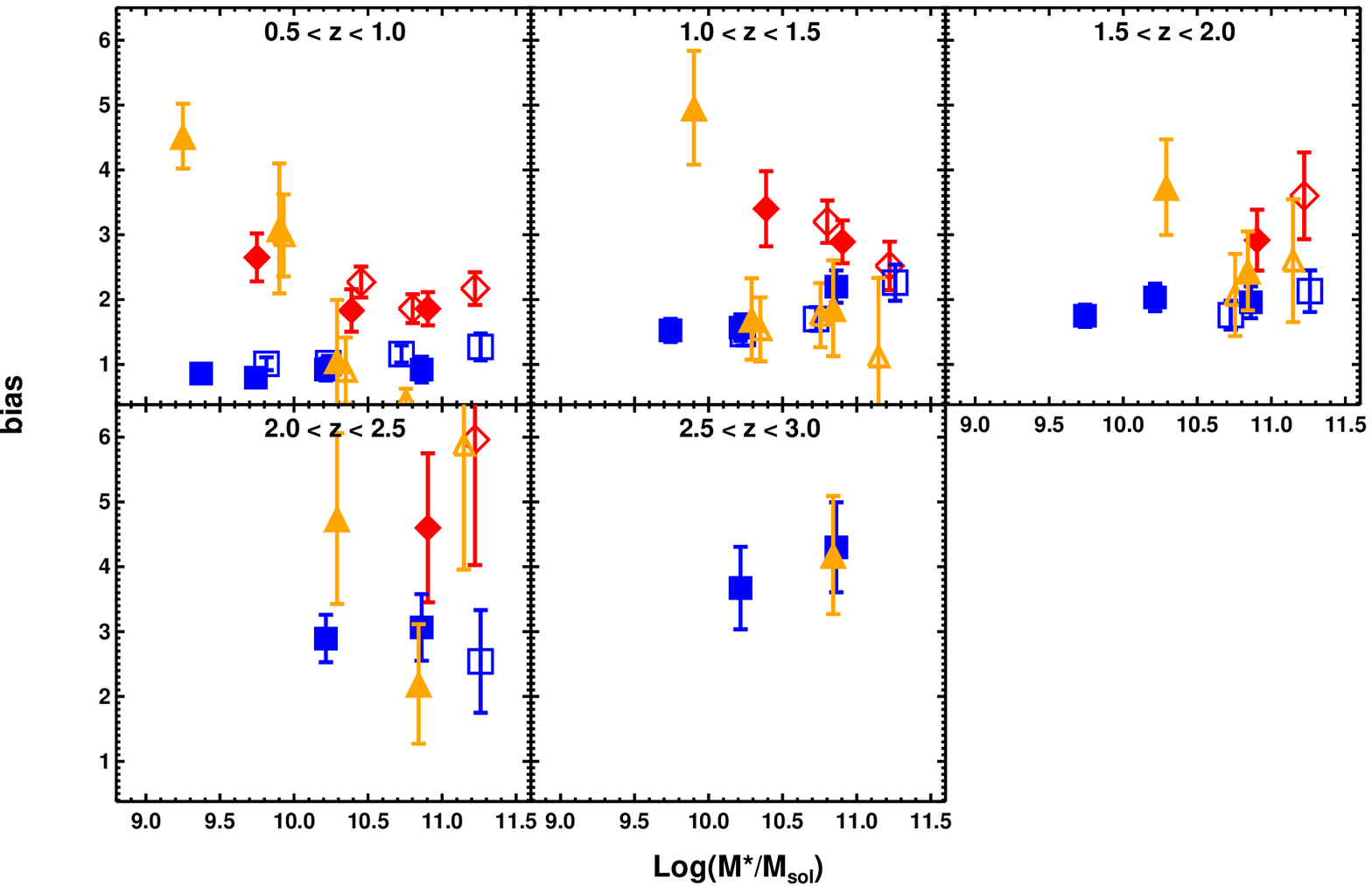}
\end{subfigure}

\begin{subfigure}
\centering
\includegraphics[height=0.6\textwidth]{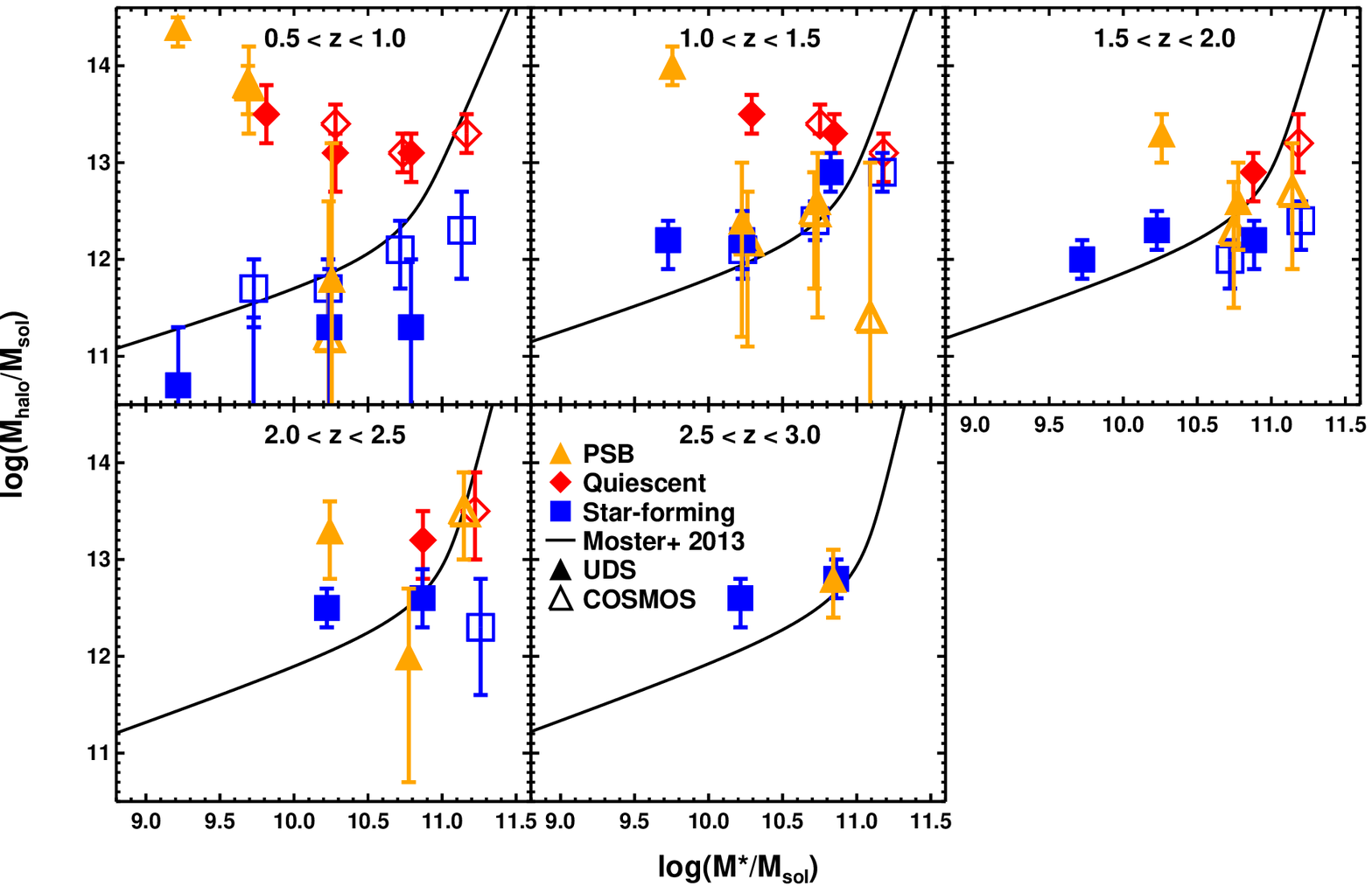}
\end{subfigure}
\caption{Top panel: The bias measurements of PSB (orange triangles), quiescent (red diamonds) and star-forming (blue squares) galaxies in the UDS (filled symbols) and COSMOS (open) fields, as a function of stellar mass, in different redshift intervals. Bottom panel: similar to the top panel, where instead we examine the inferred halo mass as a function of stellar mass for our galaxy populations. We also plot the redshift-dependent stellar-to-halo mass relations for central galaxies from \citet{Moster2013}. We show that for star-forming galaxies, their stellar mass is correlated with halo mass and they are likely to be central galaxies, irrespective of stellar mass. Low-redshift passive populations exhibit an anti-correlation between stellar mass and halo mass, such that low-mass passive galaxies occupy high mass halos and are likely to be satellite galaxies. This anti-correlation appears to be stronger for PSB galaxies.
\label{fig:halomass}}
\end{figure*}

The observed relationship between the stellar masses of star-forming galaxies and their characteristic halo masses, is consistent with the scenario that star-forming galaxies are, on average, central galaxies. However, this correlation does not rule out the presence of star-forming satellites, which have in fact been observed in galaxy groups and clusters \citep[e.g.][]{Muzzin2014}. Clustering signals from star-forming satellites, which represents a small fraction of the entire star-forming population, are likely to be washed out by the clustering amplitude exhibited by the majority of star-forming galaxies. We return to the variation of clustering within the star-forming population in Section 5.4.

In agreement with previous studies \citep[e.g.][]{Hartley2013}, passive galaxies exhibit an anti-correlation between their stellar masses and host halo masses, such that low-mass passive galaxies occupy high-mass halos. The high halo masses of low-mass quiescent and PSB galaxies imply larger satellite fractions. The halo masses of red-sequence and star-forming galaxies also appear to reflect the pivot mass scale described earlier, where the former population typically occupies halos above this mass, M$_{\text{halo}}>10^{12.5}$M$_{\odot}$. The transition population of PSB galaxies appear to disregard this halo mass scale: while low-mass PSB galaxies are in halos of mass above this limit, we find high-mass recently quenched galaxies in low-mass halos. This suggests that while most galaxies in high-mass halos are quiescent, quenched galaxies do not $necessarily$ occupy high-mass halos. A possible scenario for these galaxies, as their host halos are not massive enough (and therefore do not have enough hot gas) to keep these galaxies quenched, is that they may undergo rejuvenation episodes in the future before achieving final quiescence, as is found for high-mass PSB galaxies at $z\sim0$ \citep{Pawlik2018}.

\subsection{Rapid quenching: internal or environmentally dominated?}

We now return to a discussion of the processes driving the evolution of galaxies going through a post-starburst phase. We have seen that PSB galaxies occupy halos of varying mass, that appear to be dependent on both stellar mass and redshift. The observations of low-mass PSB galaxies in high-mass halos and high-mass PSB galaxies in low-mass halos suggest that two mechanisms could be responsible for the rapid quenching of galaxies: secular-quenching and environmental-quenching. The phenomenological formalism of \citet{Peng2010} suggests that both quenching channels are significant and independent of one another. Galaxies in the highest density environments, regardless of stellar mass, are likely to be quiescent, as are the most massive galaxies, irrespective of the environment that they inhabit. \citet{Peng2012} also showed that the environmental quenching is mostly confined to satellite galaxies. Our results have shown that rapidly quenched low-mass galaxies reside in high-mass halos, indicating that it is the halo or environment responsible for the rapid quenching in these galaxies. Once star-forming over sustained periods of time, PSB galaxies become rapidly quenched as they fall into cluster-like environments, becoming satellites in high-mass halos.

The presence of high-mass PSB galaxies in relatively low-mass halos portrays the importance of the secular-quenching route. These galaxies are likely to have been formed rapidly from the dissipational collapse of gas clouds, or rapid merging of many smaller galaxies \citep{Hopkins2009, Wellons2015}. These events could induce intense star-formation for a short period of time before being rapidly quenched due to feedback mechanisms, which may be some combination of AGN and stellar processes.

Observations of the stellar mass functions presented in \citet{Wild2016}, the morphologies \citep{Almaini2017,Maltby2018} and the inference of the clustering measurements suggest that both of the two mechanisms described above are significant. However, the contribution from each mechanism to rapid quenching of galaxies appears to change with cosmic time. The prevalence of high-mass PSB galaxies in low-mass halos at high redshifts, the compact spheroidal nature of these galaxies \citep{Almaini2017,Maltby2018}, and the lack of low-mass PSB galaxies inferred from the UDS mass functions \citep{Wild2016}, implies that most of the rapid quenching takes place in high-mass galaxies, i.e., the secular-quenching channel is the most dominant mechanism at high redshifts ($z>1$). As the Universe ages, the stellar mass distribution of the PSB galaxies shifts to lower masses, while being quenched in increasingly high-mass halos. These low-mass galaxies are also more disk-like, bearing resemblances of the star-forming progenitors from which they transition. Hence, the environmental quenching appears to become the most dominant quenching mechanism at low redshifts ($z<1$), for rapidly quenched galaxies.

\subsection{Are PSB galaxies descendants of submillimetre galaxies?}

The photometric identification of PSB galaxies is sensitive only to the strongest PSB spectral features. To produce these spectral features, these galaxies must have gone through a brief period of extremely luminous star formation before being rapidly quenched. It is natural, therefore, to explore the possible link between $z>1$ PSB galaxies and submillimetre galaxies \citep[SMGs, e.g.][]{Smail1997,Barger1998,Hughes1998}, the latter being the most luminous star-forming galaxies in the Universe. A number of properties of these two populations are highly comparable: similar characteristic stellar masses of log(M$^*$/M$_{\odot})\sim10.4-10.5$ \citep{Simpson2014, Wild2016} and PSB galaxies having space densities several times higher at $\sim6-7\times10^{-5}$ Mpc$^{-3}$, but an observable time-scale up to an order of magnitude longer \citep[100\,Myr for SMGs, a maximum visibility timescale of 1\,Gyr for PSB galaxies][]{Chapman2005, Swinbank2006, Hainline2011,Wild2016, Wild2020}.

The SFHs of PSB galaxies, as measured by \citet{Wild2020}, suggest that they have peak SFRs consistent with those derived for SMGs \citep[20-1030\,M$_{\odot}$yr$^{-1}$,][]{Swinbank2014}, which affirms their association. The two populations are also similar in their sizes. The most luminous regions of SMGs are highly compact and disky in the far-infrared \citep[e.g.,][]{Simpson2015a, Ikarashi2015, Hodge2016, Ikarashi2017}, indicating highly nucleated dust-obscured star formation. Massive PSB galaxies are compact, but they are also spheroidal in nature, with high Sersic indices \citep{Almaini2017,Maltby2018}. The possible evolutionary connection between the two populations requires dramatic structural transformation during the SMG phase, before leaving behind a quiescent ultra-compact spheroid. AGN or stellar feedback may be invoked to remove the remaining outer-lying gas that would be required for ongoing star formation.

The connection between the SMG and PSB populations is enhanced when we consider the clustering measurements. According to \cite{Wilkinson2017}, SMG activity shifts from high-mass halos (M$_{\text{halo}}>10^{13}$\,M$_{\odot}$) at $z>2.5$, to low-mass halos (M$_{\text{halo}}\sim 10^{11}$\,M$_{\odot}$) at $1<z<2$ - undergoing halo downsizing as the Universe ages. In Figure 7, we plot the predicted galaxy bias evolution of these SMGs as presented in \cite{Wilkinson2017}, calculated using the \citet{Fakhouri2010} formalism of the median growth rate of halos, as a function of halo mass and redshift. We then recompute the bias measurements for log(M$^*$/M$_{\odot})>10.2$ PSB galaxies in the UDS (which corresponds to the $90\%$ mass completeness limit for PSBs at $z=3$). From the clustering measurements in Figure 7, it is apparent that after evolving SMG halo masses for $\sim1$\,Gyr (as indicated by the green crosses in Figure 7), the resultant halos are consistent with those hosting high-mass PSB galaxies. This close correspondence in clustering, in addition to their stellar masses, SFHs and sizes, emphasises the likely close connection between the two populations.

\begin{figure}
\centering
\includegraphics[height=0.37\textwidth]{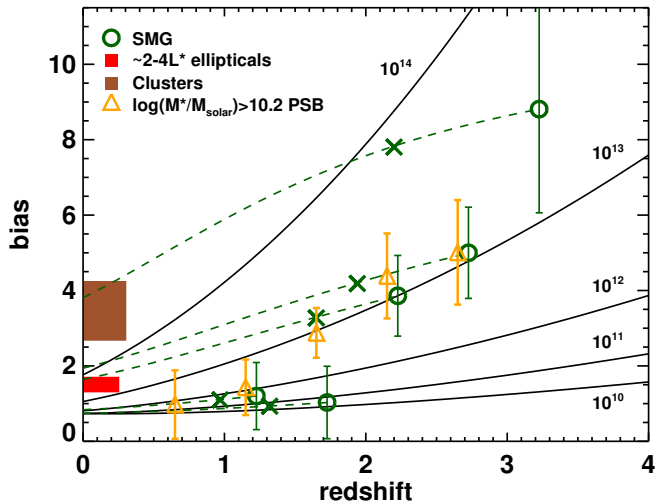}
\caption{Predicted galaxy bias evolution of SMGs (open green circles) at $z>1$,
plotted against redshift, from \citet{Wilkinson2017}. The bias values for log(M$^*/$M$_{\odot})>10.2$ PSB galaxies in UDS are plotted in open orange triangles. The solid black lines represent the expected bias evolution for dark matter halos with constant masses (labelled, in solar masses).
The mass growth of halos hosting SMGs is traced by green dashed lines, using
the formalism from \citet{Fakhouri2010}. The green crosses indicate this halo mass growth after evolving for 1\,Gyr. Typical bias measurements of $\sim2-4$\,L$^*$ galaxies (from the luminosity-bias relation derived in \citealt{Zehavi2011})
and optically selected galaxy clusters at redshifts $0.1<z<0.3$ \citep{Estrada2009} are also plotted. Consistent clustering measurements reveal that SMGs are likely to undergo a PSB phase before reaching quiescence.
\label{fig:halomass}}
\end{figure}

\subsection{The dependence of M$^{*}$ and SFR on clustering}

We now explore the clustering of galaxies in the M$^{*}$-SFR plane. Due to the sample sizes available, we study the COSMOS and UDS galaxy samples in the redshift interval $0.5 < z < 1.0$, retaining quiescent and star-forming galaxies above their respective $90\%$ mass completeness limits at $z = 1.0$. Within the quiescent and star-forming classes, we split the galaxy samples in the M$^*$-SFR plane using Voronoi binning to sort galaxies into bins of approximately equal sample size\footnote{We utilise the publicly available \texttt{voronoi\_2d\_binning} package \citep{Cappellari2003}.}. A sample size of $\sim600$ is chosen, though the results are robust to changes in binning method and sample size. Due to low sample sizes, we group together all PSB galaxies with M$^{*}>$M$_{\text{lim}, z=1}$, the mass completeness limit at $z=1$, into one bin.

\begin{figure}
\centering
\includegraphics[height=0.38\textwidth]{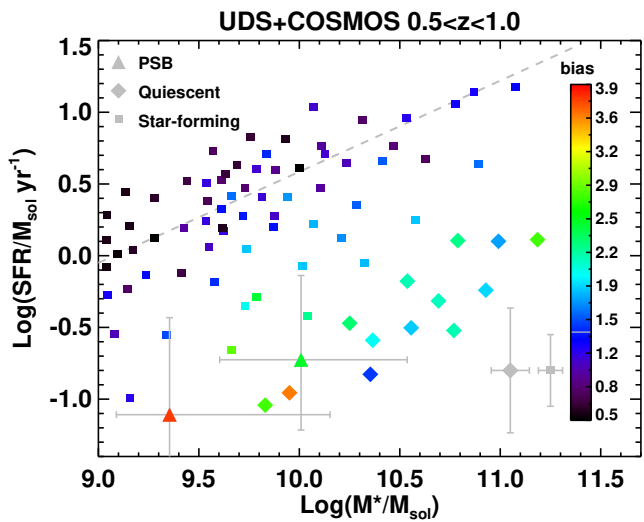}
\caption{$Top~panel$: The colour-coded bias values as a function of stellar mass and SFR, of $0.5<z<1.0$ quiescent (diamonds), star-forming (squares) and PSB (triangles) galaxies, where the former two populations are separated into Voronoi bins in the M$^*$-SFR plane, each with sample size $n\sim600$. The binning and clustering analysis were computed for the UDS and COSMOS fields separately, but the results from both fields are plotted in the same figure. The $16^{th}$ and $84^{th}$ precentiles of the PSB stellar mass and SFR distributions are indicated by solid grey bars. The grey points and their error bars located at the bottom right indicate the typical 1\,$\sigma$ errors on the M$^*$-SFR centroids for the star-forming and quiescent samples. The star-forming main sequence at $z\sim0.75$, determined by \citet{Whitaker2012b}, is plotted with a grey dashed line. Galaxies further away from the main sequence in the M$^*$-SFR plane are more strongly clustered.
\label{fig:halomass}}
\end{figure}

For each of the Voronoi bins, we derive bias measurements via a cross-correlation clustering analysis as described in Section 3. We present the results in Figure 8, by plotting the centroids of each binned sample in the M$^*$-SFR plane, colour-coded by the bias measurements. Since the bias measurements for the UDS and COSMOS fields are consistent, regardless of the position in the M$^*$-SFR plane, the results for the two fields are plotted in the same figure. The star-forming main sequence as determined by \citet{Whitaker2012b} is also included. In Figure 9, we plot the bias measurements as a function of the stellar mass, SFR and sSFR centroids.

\begin{figure*}
\centering
\includegraphics[height=0.39\textwidth]{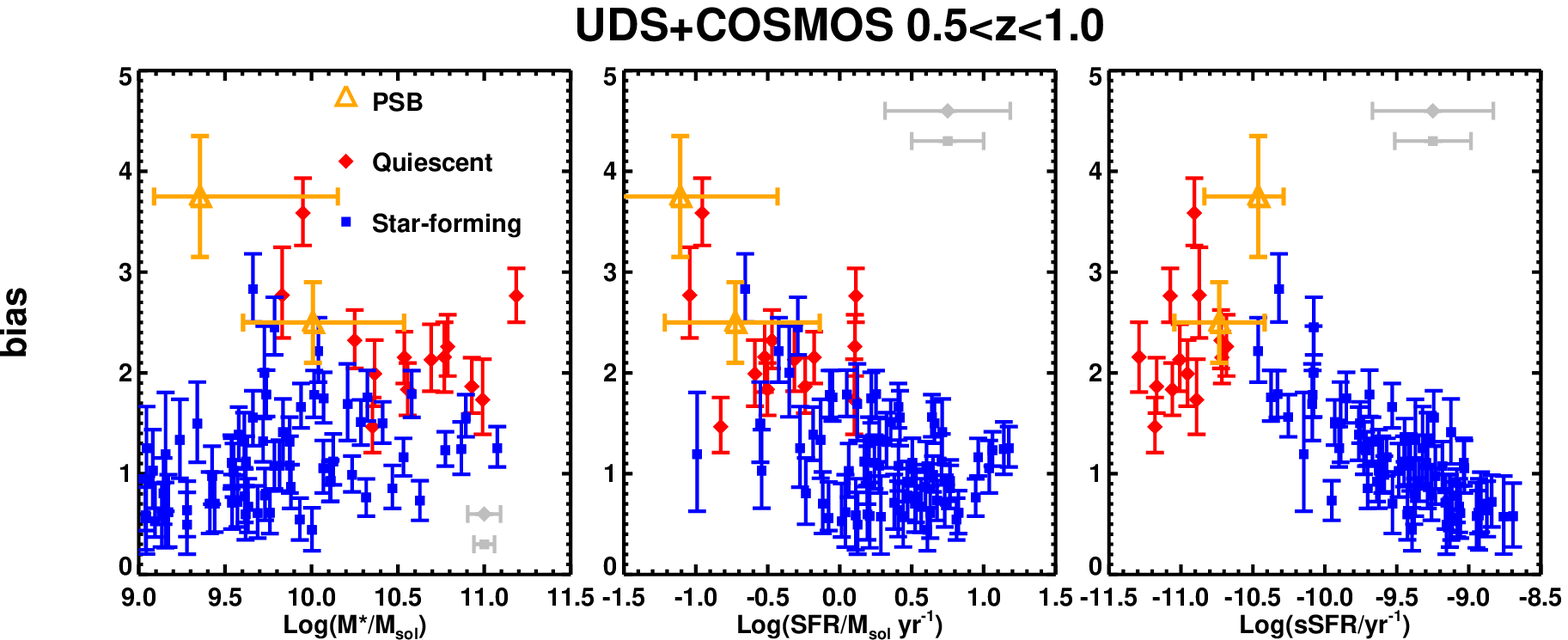}
\caption{The clustering measurements of the UDS and COSMOS $0.5<z<1.0$ galaxies, as a function of (from left to right) log stellar mass (in solar masses), log of star-formation rate and the log of the specific star-formation rate. Galaxies are also divided into various populations: red-sequence (red diamonds), star-forming (blue crosses) and PSB galaxies (orange triangles), with the former two populations further separated into Voronoi bins of sample size $n\sim600$. The $16^{th}$ and $84^{th}$ precentiles of the PSB stellar mass, SFR and sSFR distributions are indicated by their horizontal bars. The grey points and their horizontal error bars indicate the typical 1\,$\sigma$ uncertainties on the M$^*$, SFR and sSFR centroids for the star-forming and quiescent samples. The clustering measurements are highly anti-correlated with the sSFR of the galaxy, suggesting that the SFR of a galaxy and the mass of its halo are intimately related at this epoch.
\label{fig:halomass}}
\end{figure*}

From Figure 8, a number of trends are apparent. Although there is scatter in the clustering measurements along the star-forming main sequence, we find that the increase in the clustering amplitude is associated with the deviation from the main sequence in the M$^*$-SFR plane. Star-forming galaxies below the main sequence appear to be more strongly clustered. The continuous trend in bias measurements from the star-forming to the red-sequence region indicates a steady progression rather than two completely independent populations. For the passive sequence, those with the lowest stellar masses are the most clustered. As discussed in the previous section, these galaxies are more likely to be low-mass satellites being quenched in high-mass halos.

The clustering measurements in Figure 9 indicate a weak relationship with stellar mass, particularly for the star-forming class: generally, more massive galaxies are strongly clustered and inhabit higher mass halos, consistent with the findings of previous studies \citep[e.g.,][]{Wake2011, McCracken2015, Skibba2015}. However, the relationship between the galaxy stellar mass and the halo mass appears to be more complex, since we find both low-mass recently quenched galaxies and massive quiescent galaxies in high-mass halos. Moreover, this mass trend appears to be a manifestation of the passive/star-forming state of the galaxies: the relationship between the host halo mass and the stellar mass of the inhabiting galaxies is more likely to be caused by the increasing fraction of passive galaxies with higher stellar mass \citep{Hartley2013}. These galaxies may have been formed at earlier epochs in massive halos, and accumulated their mass from a combination of accretion/gas-rich mergers before quenching, and dry mergers since quenching. When separating out the galaxy populations, the correlation between the bias and stellar mass does not appear to be as apparent (when compared to trends with SFR).

The relationship between the bias measurements, and the galaxy SFRs and sSFRs suggest that high-mass halos host galaxies with lower star formation rates, whereas low-mass halos promote star formation. These trends are observed within the star-forming populations, indicating that the relationship between halo mass and SFR is continuous. Star-forming galaxies with relatively low SFRs/sSFRs (SFR $\sim1$\,M$_{\odot}$\,yr$^{-1}$, sSFR $\sim10^{-10}$\,yr$^{-1}$) inhabit the region of SC1-SC2 close to the red-sequence boundary, indicating that these may be `green valley' galaxies in high-mass halos that are about to enter, or are in the early stages of, a quenching phase. Whether this is rapid or slow remains to be seen, however, though the proximity of these galaxies to the PSB region in super-colour space may indicate the quenching may eventually be rapid.

The presence of these strongly clustered star-forming galaxies (in halos of mass $\sim10^{13}$M$_{\odot}$) indicate that the relationship between the halo mass and the activity of a galaxy is more complex than previously considered. We discussed earlier the notion of a halo mass scale \citep[$\sim10^{12.5}$M$_{\odot}$; e.g.][]{Croton2006,Cen2011}, above which galaxies start to become quenched, indicating there are two modes of clustering. However, the results suggest a more continuous relationship between halo mass and the galaxy's SFR, in which the halo mass regulates the rate at which inhabiting galaxies can form new stars. The fact that we find low-mass $rapidly$-quenched galaxies in high-mass halos imply that the regulation of star formation in low-mass galaxies is most efficient in high-mass halos.

\subsection{Cosmic Variance}
Clustering analyses in pencil-beam surveys such as the UDS and COSMOS are likely to be affected by the density fluctuations in the large-scale structure. This so-called cosmic variance becomes more significant for massive high-redshift galaxies. We therefore evaluate the impact cosmic variances may have on our clustering measurments, by using the recipe of \citet{Moster2011}. Estimating the cosmic variance, $\sigma_{cv}$, in every redshift and stellar mass employed for our work, we find $\sigma_{cv}$ to vary from 0.067 to 0.141, and 0.060 to 0.135, for the UDS and COSMOS surveys, respectively. The aforementioned upper limits of $\sigma_{cv}$ correspond to the most massive and highest redshift intervals. Compared to the size of the errors derived from the clustering analysis (particularly for the PSB and quiescent populations, where $\sigma_b \sim 0.5 - 1$), the uncertainties from cosmic variance produce limited impact on our clustering measurements and conclusions. Hence, our biggest source of error is likely to be the limited size of our samples and it is only by measuring the clustering over a larger survey area that we will achieve smaller uncertainties of our measurements. Nevertheless, the errors from cosmic variance are added in quadrature to the uncertainties of the bias measurements and all figures and tables presented in this work include the addition of this cosmic variance.

We also remark on the impact the $z \sim 0.65$ supercluster in the UDS field may have on our clustering measurements, particularly those of the PSB galaxy population. It is possible that this supercluster may be biasing our clustering measurements, so that the low-mass PSB inferred halo mass measurements (M$_{\text{halo}}>10^{14}$M$_{\odot}$) are no longer representative and comparable to that of the COSMOS field. This supercluster is indeed one of 11 cluster candidates at redshifts $0.5 < z < 1.0$ that contain at least 45 ``Friends of friends'' members, although their cluster/halo masses are not known \citep{Socolovsky2018}. Similarly, the COSMOS field hosts 8 massive (M$_{200}>10^{14}$M$_{\odot}$) clusters at redshifts $0.5<z<1.0$, detected with XMM-Newton \citep{Finoguenov2007}. Taking into account the differences in how these clusters were identified, their prevalence is consistent with the numbers predicted by the halo mass function of \citet{Tinker2008}, that is, $\sim6$ and $\sim9$ massive (M$_{\text{halo}}>10^{14}$M$_{\odot}$) halos (at redshifts $0.5<z<1.0$) for the UDS and COSMOS fields, respectively. We are therefore confident that the large bias inferred for the lowest-mass low-redshift PSB galaxies in the UDS field is part of the anti-correlation relationship between stellar mass and bias for quiescent and PSB galaxies, and is not significantly affected by the known supercluster feature.


\section{CONCLUSIONS}
\label{sec:Conclusion}
We have used a cross-correlation analysis to study the clustering and halo masses of PSB galaxies to $z=3.0$. These galaxies were selected photometrically in the UDS and COSMOS using a PCA technique, producing a sample of $\sim4000$ PSB galaxies, the largest high-redshift sample to date. We cross-correlated PSB galaxies with a much larger sample of $K$-selected tracer galaxies, providing reasonably well constrained halo masses for the first time. We interpreted how the clustering measurements of PSB galaxies compare to those of the star-forming and quiescent galaxies and in particular, the mechanisms that drive the evolution of these populations to present day. Our main results are summarised as follows.

\begin{enumerate}
\item PSB galaxies occupy dark matter halos of varying mass, depending on redshift and stellar mass. In particular, low-mass, low-redshift ($0.5<z<1.0$) PSB galaxies occupy very high-mass halos, with M$_{\text{halo}}>10^{14}$\,M$_{\odot}$, consistent with them being low-mass satellites quenched in dense environments. There is tentative evidence that high-mass ($>10^{10}$M$_{\odot}$) PSB galaxies undergo halo downsizing, occupying lower mass halos as the Universe ages. These results suggest that both secular-quenching and environmental-quenching mechanisms are responsible. Secular-quenching is likely to be the dominant driver of rapid quenching at high redshifts $z>1.5$, where the progenitors of PSB galaxies may undergo rapid dissipational collapse and gas-rich mergers. At $z<1$, the key driver of rapid quenching, particularly in low-mass galaxies, appears to be driven by processes associated with dense environments or high-mass halos.
\item The clustering measurements and the apparent halo downsizing of SMGs and PSB galaxies are consistent with the scenario in which SMGs are the progenitors of high-mass PSB galaxies.
\item We present the clustering analysis on galaxies in the M$^*$-SFR plane. Sorting the galaxies in this plane using a Voronoi binning technique and performing clustering analyses with these subpopulations, we find that the clustering strengths of galaxies are correlated with their SFRs, such that galaxies with lower SFRs are more strongly clustered and reside in higher mass halos. The continuous trend between halo mass and SFR is consistent with the scenario in which the mass of the dark matter halos regulates the rate at which galaxies can form new stars.
\end{enumerate}

The clustering analysis has revealed the first insights into the properties of halos hosting PSB galaxies. However, we still find limitations on this analysis due to the small samples of this rare population to date. The next generation of optical/NIR surveys, such as Euclid - which will survey $>10^9$ galaxies across 15,000 square degrees of the sky \citep{Laureijs2011} - will offer the unprecedented statistical power needed to constrain the halo masses of galaxy populations and to augment our understanding of the galaxy-halo connection.


\section{Data availability}
A public release of the processed data and catalogues of UDS-DR11 forming part of the basis of this work is in preparation. Details can be obtained from Omar Almaini (omar.almaini@nottingham.ac.uk). In the meantime, data will be shared on request to the corresponding author. The COSMOS2015 catalogues are publically available at \url{https://ftp.iap.fr/pub/from_users/hjmcc/COSMOS2015/} and details can be found in the data release paper of \citep{Laigle2016}.

\section{Acknowledgements}
AW and VW acknowledges support from the European Research Council Starting Grant (SEDMorph; P.I. V.~Wild). AW also acknowledges funding from the STFC and the H2020 ERC Consolidator Grant 683184. This work is based in part on observations from ESO telescopes at the Paranal Observatory (programmes 180.A-0776, 094.A-0410,  and 194.A-2003). AW wishes to thank Kristin Coppin, James Bolton, Ryan Hickox, Arjen van der Wel, Charutha Krishnan, Miguel Socolovsky and Elizabeth Cooke for many useful discussions. We extend our gratitude to the staff at UKIRT for their tireless efforts in ensuring the success of the UDS project. We also wish to recognize and acknowledge the very significant cultural role and reverence that the summit of Maunakea has always had within the indigenous Hawaiian community. We were most fortunate to have the opportunity to conduct observations from this mountain. Finally, we thank the anonymous referee for their helpful feedback, which led to the improvement in the quality of this manuscript.

\begin{table*}
\centering
\begin{tabular}{ c c c c c c c c }
\hline
$z$ & log(M$^{*}$/M$_{\odot}$) & N$_{\text{gal}}$ & $\Sigma_{\text{weight}}$ & $b$ & $\sigma_{b}$ & log(M$_{\text{halo}}$/M$_{\odot}$) & $\sigma_{\text{M}_{\text{halo}}}$\\
\hline
\textbf{post-starburst} \\
$[0.5,1.0]$ & $[9.0,9.5]$ & 234 & 200.17 & 4.94 & 0.68 & 14.4 & $[14.2,14.5]$ \\
$[0.5,1.0]$ & $[9.5,10.0]$ & 69 & 58.64 & 3.10 & 1.00 & 13.8 & $[13.3,14.2]$ \\
$[0.5,1.0]$ & $[10.0,10.5]$ & 44 & 36.55 & 1.06 & 0.93 & 11.8 & $[8.4,13.2]$ \\
$[1.0,1.5]$ & $[9.5,10.0]$ & 179 & 112.82 & 4.96 & 0.88 & 14.0 & $[13.8,14.2]$ \\
$[1.0,1.5]$ & $[10.0,10.5]$ & 202 & 146.70 & 1.70 & 0.63 & 12.4 & $[11.2,13.0]$ \\
$[1.0,1.5]$ & $[10.5,12.0]$ & 136 & 105.27 & 1.87 & 0.74 & 12.6 & $[11.4,13.1]$ \\
$[1.5,2.0]$ & $[10.0,10.5]$ & 144 & 125.05 & 3.73 & 0.74 & 13.3 & $[13.0,13.5]$ \\
$[1.5,2.0]$ & $[10.5,12.0]$ & 178 & 158.10 & 2.44 & 0.61 & 12.6 & $[12.1,13.0]$ \\
$[2.0,2.5]$ & $[10.0,10.5]$ & 167 & 84.64 & 4.74 & 1.32 & 13.3 & $[12.8,13.6]$ \\
$[2.0,2.5]$ & $[10.5,12.0]$ & 154 & 97.38 & 2.19 & 0.92 & 12.0 & $[10.7,12.7]$ \\
$[2.5,3.0]$ & $[10.5,12.0]$ & 150 & 119.64 & 4.18 & 0.91 & 12.8 & $[12.4,13.1]$ \\

\hline
\textbf{quiescent} \\
$[0.5,1.0]$ & $[9.5,10.0]$ & 270 & 238.77 & 2.46 & 0.46 & 13.5 & $[13.2,13.8]$ \\
$[0.5,1.0]$ & $[10.0,10.5]$ & 722 & 637.47 & 1.83 & 0.33 & 13.1 & $[12.7,13.3]$ \\
$[0.5,1.0]$ & $[10.5,12.0]$ & 1157 & 1028.61 & 1.86 & 0.26 & 13.1 & $[12.8,13.3]$ \\
$[1.0,1.5]$ & $[10.0,10.5]$ & 481 & 349.10 & 3.40 & 0.58 & 13.5 & $[13.3,13.7]$ \\
$[1.0,1.5]$ & $[10.5,12.0]$ & 1341 & 1094.49 & 2.89 & 0.33 & 13.3 & $[13.1,13.5]$ \\
$[1.5,2.0]$ & $[10.5,12.0]$ & 601 & 504.16 & 2.92 & 0.47 & 12.9 & $[12.6,13.1]$ \\
$[2.0,2.5]$ & $[10.5,12.0]$ & 252 & 139.01 & 4.60 & 1.15 & 13.2 & $[12.8,13.5]$ \\

\hline
\textbf{star-forming} \\
$[0.5,1.0]$ & $[9.0,9.5]$ & 7770 & 6315.34 & 0.86 & 0.14 & 10.7 & $[6.9,11.3]$ \\
$[0.5,1.0]$ & $[9.5,10.0]$ & 3772 & 3241.72 & 0.80 & 0.14 & 4.3 & $[8.4,11.4]$ \\
$[0.5,1.0]$ & $[10.0,10.5]$ & 2408 & 2098.53 & 0.92 & 0.17 & 11.3 & $[9.8,11.9]$ \\
$[0.5,1.0]$ & $[10.5,12.0]$ & 1860 & 1625.72 & 0.92 & 0.20 & 11.3 & $[8.4,12.0]$ \\
$[1.0,1.5]$ & $[9.5,10.0]$ & 7006 & 5049.07 & 1.52 & 0.18 & 12.2 & $[11.9,12.4]$ \\
$[1.0,1.5]$ & $[10.0,10.5]$ & 4189 & 3061.54 & 1.57 & 0.20 & 12.2 & $[11.9,12.5]$ \\
$[1.0,1.5]$ & $[10.5,12.0]$ & 3465 & 2576.52 & 2.20 & 0.25 & 12.9 & $[12.7,13.1]$ \\
$[1.5,2.0]$ & $[9.5,10.0]$ & 6226 & 4299.61 & 1.75 & 0.17 & 12.0 & $[11.8,12.2]$ \\
$[1.5,2.0]$ & $[10.0,10.5]$ & 3624 & 2647.32 & 2.03 & 0.21 & 12.3 & $[12.1,12.5]$ \\
$[1.5,2.0]$ & $[10.5,12.0]$ & 3410 & 2446.36 & 1.96 & 0.25 & 12.2 & $[11.9,12.4]$ \\
$[2.0,2.5]$ & $[10.0,10.5]$ & 3157 & 2009.71 & 2.89 & 0.37 & 12.5 & $[12.3,12.7]$ \\
$[2.0,2.5]$ & $[10.5,12.0]$ & 2185 & 1188.17 & 3.06 & 0.51 & 12.6 & $[12.3,12.9]$ \\
$[2.5,3.0]$ & $[10.0,10.5]$ & 1526 & 1053.75 & 3.67 & 0.64 & 12.6 & $[12.3,12.8]$ \\
$[2.5,3.0]$ & $[10.5,12.0]$ & 877 & 570.40 & 4.30 & 0.69 & 12.8 & $[12.6,13.0]$ \\

\hline
\end{tabular}
\caption{Table of UDS clustering measurements. The columns shown are: redshift intervals, stellar mass intervals, number of galaxies, the sum of weights (the expected number of galaxies from the redshift probability distributions), galaxy bias, the 1$\sigma$ uncertainty on the bias, halo masses converted from the bias measurements \citep{Mo2002} and their upper and lower $1\sigma$ limits.}
\label{table:1}
\end{table*}

\begin{table*}
\centering
\begin{tabular}{ c c c c c c c c }
\hline
$z$ & log(M$^{*}$/M$_{\odot}$) & N$_{\text{gal}}$ & $\Sigma_{\text{weight}}$ & $b$ & $\sigma_{b}$ & log(M$_{\text{halo}}$/M$_{\odot}$) & $\sigma_{\text{M}_{\text{halo}}}$\\
\hline
\textbf{post-starburst} \\
$[0.5,1.0]$ & $[9.5,10.0]$ & 184 & 156.37 & 2.99 & 0.63 & 13.8 & $[13.5,14.0]$ \\
$[0.5,1.0]$ & $[10.0,10.5]$ & 107 & 88.82 & 0.90 & 0.51 & 11.2 & $[8.4,12.6]$ \\
$[0.5,1.0]$ & $[10.5,11.0]$ & 42 & 34.84 & 0.44 & 0.18 & 8.4 & $[8.4,8.4]$ \\
$[1.0,1.5]$ & $[10.0,10.5]$ & 334 & 258.10 & 1.54 & 0.49 & 12.2 & $[11.1,12.7]$ \\
$[1.0,1.5]$ & $[10.5,11.0]$ & 296 & 270.40 & 1.76 & 0.49 & 12.5 & $[11.7,12.9]$ \\
$[1.0,1.5]$ & $[11.0,13.0]$ & 63 & 57.87 & 1.12 & 1.21 & 11.4 & $[7.2,13.0]$ \\
$[1.5,2.0]$ & $[10.5,11.0]$ & 356 & 310.50 & 2.07 & 0.63 & 12.3 & $[11.5,12.8]$ \\
$[1.5,2.0]$ & $[11.0,13.0]$ & 126 & 115.40 & 2.60 & 0.95 & 12.7 & $[11.9,13.2]$ \\
$[2.0,2.5]$ & $[11.0,13.0]$ & 113 & 81.41 & 5.88 & 1.92 & 13.5 & $[13.0,13.9]$ \\
\hline
\textbf{quiescent} \\
$[0.5,1.0]$ & $[10.0,10.5]$ & 1489 & 1290.67 & 2.27 & 0.24 & 13.4 & $[13.2,13.6]$ \\
$[0.5,1.0]$ & $[10.5,11.0]$ & 2152 & 1987.57 & 1.86 & 0.22 & 13.1 & $[12.9,13.3]$ \\
$[0.5,1.0]$ & $[11.0,13.0]$ & 846 & 794.06 & 2.17 & 0.25 & 13.3 & $[13.1,13.5]$ \\
$[1.0,1.5]$ & $[10.5,11.0]$ & 1572 & 1357.31 & 3.20 & 0.33 & 13.4 & $[13.3,13.6]$ \\
$[1.0,1.5]$ & $[11.0,13.0]$ & 881 & 801.43 & 2.52 & 0.37 & 13.1 & $[12.8,13.3]$ \\
$[1.5,2.0]$ & $[11.0,13.0]$ & 526 & 456.91 & 3.60 & 0.67 & 13.2 & $[12.9,13.5]$ \\
$[2.0,2.5]$ & $[11.0,13.0]$ & 188 & 128.04 & 5.96 & 1.93 & 13.5 & $[13.0,13.9]$ \\

\hline
\textbf{star-forming} \\
$[0.5,1.0]$ & $[9.5,10.0]$ & 13016 & 11667.30 & 1.01 & 0.10 & 11.7 & $[11.3,12.0]$ \\
$[0.5,1.0]$ & $[10.0,10.5]$ & 7410 & 6647.65 & 1.02 & 0.11 & 11.7 & $[11.3,12.0]$ \\
$[0.5,1.0]$ & $[10.5,11.0]$ & 3194 & 2956.38 & 1.16 & 0.13 & 12.1 & $[11.7,12.4]$ \\
$[0.5,1.0]$ & $[11.0,13.0]$ & 512 & 486.19 & 1.27 & 0.21 & 12.3 & $[11.8,12.7]$ \\
$[1.0,1.5]$ & $[10.0,10.5]$ & 9377 & 7656.27 & 1.47 & 0.18 & 12.1 & $[11.8,12.3]$ \\
$[1.0,1.5]$ & $[10.5,11.0]$ & 4453 & 3797.09 & 1.70 & 0.18 & 12.4 & $[12.2,12.6]$ \\
$[1.0,1.5]$ & $[11.0,13.0]$ & 937 & 846.66 & 2.26 & 0.28 & 12.9 & $[12.7,13.1]$ \\
$[1.5,2.0]$ & $[10.5,11.0]$ & 4226 & 3194.83 & 1.76 & 0.22 & 12.0 & $[11.7,12.2]$ \\
$[1.5,2.0]$ & $[11.0,13.0]$ & 1004 & 830.88 & 2.13 & 0.32 & 12.4 & $[12.1,12.6]$ \\
$[2.0,2.5]$ & $[11.0,13.0]$ & 538 & 422.47 & 2.54 & 0.79 & 12.3 & $[11.6,12.8]$ \\

\hline
\end{tabular}
\caption{Table of COSMOS clustering measurements: all columns have the same meaning as Table 2.}
\label{table:1}
\end{table*}

\small
\bibliographystyle{mnras}
\bibliography{AWreferences}

\bsp

\label{lastpage}

\end{document}